\documentclass[11pt,a4paper]{article}
\usepackage{jheppub}
\usepackage[utf8]{inputenc} 
\usepackage{multirow}    
\usepackage{xcolor}
\usepackage{graphicx}
\usepackage{amsmath}
\usepackage{amssymb}
\usepackage{braket} 
\usepackage{hyperref}
\usepackage{frontespizio}
\usepackage{comment}  
\usepackage{epsfig}
\usepackage{float}
\usepackage{bbold}
\usepackage[title]{appendix}
\usepackage{multirow} 
\usepackage{tikz}
\usetikzlibrary{decorations.pathmorphing}
\usepackage{varwidth}
\usepackage{tikz}
\usetikzlibrary{backgrounds,patterns,calc}
\usepackage{xparse}  
\usepackage{subfigure}  
\usepackage{bbm}
\usepackage{wasysym}

\usepackage{jheppub} 

\title{Categorical Symmetries and Fiber Functors from Multiple Gaugeable Homomorphisms from 6D ${\cal N}=(2,0)$ SCFTs}

\author[]{Veronica Pasquarella}
\affiliation{Department of Applied Mathematics and Theoretical Physics (DAMTP),}
\affiliation{University of Cambridge,}  
\affiliation{Wilberforce Road, CB3 0WA, Cambridge, UK}

\emailAdd{vp360@damtp.cam.ac.uk}

\abstract{Exploiting the symmetry topological field theory/topological order correspondence (SymTFT/TO), together with the higher-categorical structure of 6D ${\cal N}=$(2,0) SCFTs, we prove that the total quantum dimension of the relative gaugeable algebra leading to intrinsic non-invertible symmetries between class ${\cal S}$ theories is greater with respect to the non-intrinsic case. From a higher-categorical perspective, this supports the idea that multiplicity is allowed to exceed unity in some superselection sectors.}

\makeatletter
\gdef\@fpheader{}
\makeatother
\begin{document} 
\maketitle

\section{Introduction}  

Since first encountered, \cite{Witten:1995zh}, 6D ${\cal N}=(2,0)$ SCFTs, also referred to as \emph{theory X}, have attracted great interest in, both, the mathematics and high-energy physics community alike. From the high-energy physics point of view, they are essential for deriving lower-dimensional SCFTs, \cite{Witten:2007ct,Chacaltana:2012zy,Tachikawa:2013hya,Moore}, whose features are much better understood. On the other hand, from a purely mathematical perspective, they provide an extremely interesting and rich setup where to exploit the present-day knowledge of representation theory, \cite{Witten:2009at,DBZ}.

 In recent years, the two communities have been brought together by fascinating developments on either side, \cite{Bah:2022wot}, highlighting the importance of bridging the gap in between different formalisms, with category theory playing a key role in pursuing this task. Indeed, the pioneering works of \cite{Moore:1988qv,Moore:1988ss,Verlinde:1988sn} led to the discovery of modular tensor categories in anyon gauging theory. Since then, such deep mathematical language inspired applicability to higher-dimensional QFTs, by identifying their symmetries with topological defects, \cite{Gaiotto:2014kfa, Gaiotto:2020iye}. Of particular importance to us is the richness of higher-categorical theory applied to topological orders (TOs), \cite{Kong:2020cie,Kong:2022cpy,Kong:2019byq,Kong:2019cuu,Kong:lastbutone,Gaiotto:2019xmp,Johnson-Freyd:2021tbq,Johnson-Freyd:2020usu,MYu}, and symmetry topological field theories (SymTFTs), \cite{Freed:2012bs,Freed:2022qnc,Freed:2022iao}. 

Embracing the thrust of recent developments in the field, the present work is meant to be the first of a series of papers where the author applies the techniques of category theory to furthering the understanding of 6D ${\cal N}=(2,0)$ SCFTs and theories derived from them. In doing so, we will attempt, wherever possible, to provide an understanding of the procedure as well as of the outcome of our findings, from, both, a mathematician's and a theoretical physicist's point of view. Hopefully, this will further motivate both communities to benefit form each other's influence. 

This article aims at furthering the understanding of the categorical structure arising from dimensional reduction of 6D ${\cal N}=(2,0)$ SCFTs. In particular, our aim is that of identifying a criterion to distinguish between intrinsic and non-intrinsic non-invertible symmetries\footnote{See, for example, \cite{Bashmakov:2022jtl,Bashmakov:2022uek,Bhardwaj:2022kot,Bhardwaj:2022yxj,Kaidi:2021xfk,Kaidi:2022uux,Kaidi:2022cpf,Burbano:2021loy,Choi:2022zal,Choi:2021kmx, Antinucci:2022vyk} for explanation on the current state of the art.} in terms of the quantum dimension\footnote{The total quantum dimension is defined from the braiding and fusion rules of the underlying categorical structure of the topological symmetries of the theory in question. Further explanation will be provided in due course.} of gaugeable algebras\footnote{Our criterion is compatible with the proposal put forward by \cite{Antinucci:2022cdi}, where the authors introduce the notion of the \emph{rank} of duality defects in class ${\cal S}$ theories. However, we wish to emphasise that our formulation is more suitable for, and mostly motivated by, currently ongoing work by the same author, and will soon be reported, \cite{VP}.}. In doing so, we rely upon their description in terms of relative field theories, as described in \cite{Bashmakov:2022uek} in terms of the Freed-Moore-Teleman setup, \cite{Freed:2012bs,Freed:2022qnc,Freed:2022iao}. Our main proposal is that the total quantum dimension of the gaugeable algebra implementing the gauging in the bulk Sym TFT is a key quantity to distinguish in between the two cases. Specifically, for the intrinsic case, multiplicity is greater w.r.t. the non-intrinsic case, thereby signalling the possibility for additional d.o.f. to be stored in certain superselection sectors of the resulting absolute theory. 

Our results extend arguments proposed in \cite{Kaidi:2022cpf}, where the authors also proposed a way of distinguishing intrinsic from non-intrinsic non-invertibility in 2D by means of the quantum dimension of the non-invertible defect. 

The present work is organised as follows: section \ref{sec:BMTCs} provides a brief overview of some key features of higher-categorical structures that are relevant for our analysis. In the first part of this section, we briefly recall the relation between the total quantum dimension, fusion and braiding structures. As an example of a higher-categorical structure, we describe the Chern-Simons theory for a flux-charge-quasiparticle setup, and how  gauging a subgroup of the anyonic symmetry affects the total quantum dimension. 

Section \ref{sec:Freed-Moore-Teleman} opens with an explanation for how to gauge a higher-categorical structure by means of algebraic gauging, \cite{TJF,Kong:2013aya,Yu:2021zmu}, emphasising the richer structure resulting from the case in which gauging of multiple subalgebras takes place. We then turn to implementing such techniques to theories descending from 6D ${\cal N}=(2,0)$ SCFTs upon dimensional reduction over a Riemann surface. At first, we briefly recall the realisation of 6D theories with maximum amount of supersymmetry as a relative QFT in the Freed-Teleman setup, and how to recover intrinsic non-invertible symmetries in class ${\cal S}$ theories by performing a double flooding of the Sym TFT according to the SymTFT/TO correspondence. We show that the quantum dimension of the relative gauging algebra interpolating between different absolute 4D gauged theories in the intrinsic case always exceeds that featuring in the non-intrinsic case. We thereby propose this as a general criterion to be applied to class ${\cal S}$ theories descending from 6D ${\cal N}=(2,0)$ SCFTs. En-passing, we comment on a proposal regarding the realisation of such composite absolute theories and the definition of a fiber functor intrinsically related to the notion of a partition function for a 3D theory rather than a 4D theory, which is current work in progress by the same author, \cite{VP}.

 In an upcoming work \cite{VP}, we reported a continuation of this analysis, by making use of tools such as factorisation homology and full-dualisability in the context of cobordism theory, with suitable adaptation to class-${\cal S}$ by means of the AGT correspondence. We expect further investigation in terms of cobordisms could lead to potential further understanding of singularity theories and moduli spaces of varieties arising in algebraic geometry settings, which we aim to address in future work.

\section{Higher-categorical structures }   \label{sec:BMTCs}

For a QFT to be well defined, its spectrum of allowed operators must be compatible with the symmetries of the given theory, \cite{Aharony:2013hda}. This statement has been a very active area of research, and mostly motivated furthering a full mathematical formulation of QFTs. Category theory has been pointed out as a promising candidate in pursuing such aim. 

Following suite, the present work aims at exploring the higher-categorical theory of and descending from 6D ${\cal N}=(2,0)$ SCFTs with the aim of identifying a categorical theory quantity enabling to distinguish between intrinsic and non-intrinsic non-invertible symmetries in class ${\cal S}$ theories obtained by dimensional reduction of their parent theory. 

Before delving into the specifics of the 6D theory, to which section \ref{sec:Freed-Moore-Teleman}    is devoted, the current section aims at providing a brief overview of some essential tools we will need in the remainder of the paper, as well as further motivating the use of such mathematical tools within the following theoretical physics treatment. This section is structured as follows:

\begin{itemize}  

\item In the first part of this section we briefly overview some key definitions of higher-categories. 

\item As a specific example, we outline the flux-charge-quasiparticle treatment in terms of Chern-Simons (CS) theory, and how this is related to representation theory.   

\end{itemize}

\subsection{Defect fusion categories}   \label{sec:dfc}

\emph{Categories} are defined as sets of objects related to each other by morphisms. An $n$-category is a collection of categories related to each other by \emph{functors}. Groups constitute a special type of categories, where only one element is present, and every morphism is an isomorphism. A \emph{higher}-category, or $n$-category, is an $(\infty,1)$-category enriched with $(n-1)$-categories, with $n$ denoting the dimensionality.

A category, ${\cal C}$, is \emph{monoidal} (MC) if it comes equipped with a bi-functor   

\begin{equation}    
\otimes\ :\ {\cal C}\times{\cal C}\ \rightarrow\ {\cal C}.
\label{eq:mult}    
\end{equation}  

For ${\cal C}$ semisimple, the tensor product of any two simple objects $x,y\in{\cal C}$ is 

\begin{equation}  
x\ \otimes\ y\ \simeq\ \bigoplus_{z\in\text{Irr}({\cal C})}\ N_{_{xy}}^{^{z}}\cdot z     
\ \ \ \ 
,\ \ \ \ N_{_{xy}}^{^{z}}\in\mathbb{N},
\end{equation}  
where $\left\{N_{_{xy}}^{^{z}}\right\}_{_{x,y,z\in\text{Irr}({\cal C})}}$ denote the \emph{fusion rules} of ${\cal C}$. Equivalently, the isomorphism classes of simple objects ${\cal C}$ generate the \emph{fusion ring} or \emph{Grothendieck ring} of ${\cal C}$, where multiplication is given by \eqref{eq:mult}, and $N_{_{xy}}^{^{z}}$ as structure constants.

Given $x, y\in\ {\cal C}$ 

\begin{equation}   
\text{Hom}:\ x,y\ \ \rightarrow\ \text{Hom}(x,y)   
\end{equation}  
is an $(n-1)$-category, and for $\mathbb{1}\in{\cal C}$  

\begin{equation}   
\text{Hom}:\ \mathbb{1}, \mathbb{1}\ \ \rightarrow\ \text{Hom}(\mathbb{1}\ ,\mathbb{1}) \ \overset{def.}{=}\ \Omega\ {\cal C},        
\end{equation}  
$\Omega\ {\cal C}$ is a monoidal $(n-1)$-category.

\medskip  

\medskip

\subsection*{Braiding and ribbon structure}

\medskip  

\medskip 

A modular category (MC) is \emph{braided} if, in addition to \eqref{eq:mult} it is also equipped with a natural isomorphism 

\begin{equation}  
c_{_{x,y}}: x\otimes y\ \rightarrow\ y\otimes x,  
\label{eq:braiding}   
\end{equation} 
called \emph{braiding}. Given two simple objects $x,y\in{\cal C}$, the \emph{S-matrix} is defined as

\begin{equation} 
\boxed{\ \ S\ \overset{def.}{=}\ \left(S_{_{xy}}\right)_{_{z\in\text{Irr}({\cal C})}},  \color{white}\bigg]\ \ }     
\label{eq:SM1}
\end{equation} 
with components defined from \eqref{eq:braiding}

\begin{equation} 
S{_{xy}}\ \overset{def.}{=} \ \text{Tr}\ \left(c_{_{y,x^{*}}}\ \circ\ c_{_{x^{*},y}}\right)\ \equiv\ \text{Tr}\ \left(c_{_{y^{*},x}}\ \circ\ c_{_{x,y^{*}}}\right).
\label{eq:SM}
\end{equation}

\begin{figure}[ht!]  
\begin{center}   
\includegraphics[scale=1]{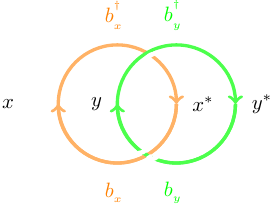}  
\ \ \ \ \ \ \ \ \ \ \ \ \ \ \ \ \ \ \ \ 
\includegraphics[scale=1]{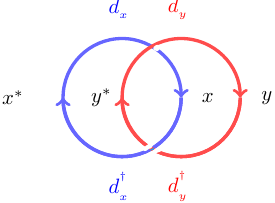}  
\caption{\small The S-matrix corresponds to the Hopf link, and encodes the information of mutual-statistics. As shown in these pictures, the linking between particle worldlines encodes information of mutual-statistics.}   
\label{fig:SM}  
\end{center}  
\end{figure}

\begin{figure}[ht!]  
\begin{center}   
\includegraphics[scale=1]{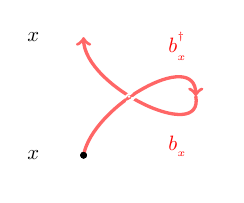}  
\ \ \ \ \ \ \ \ \ \ \ \ \ \ \ \ \ \ \ \ 
\includegraphics[scale=1]{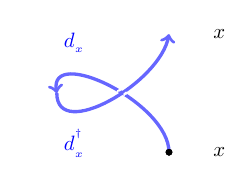}  
\caption{\small The ribbon structure describing self-statistics is here drawn as the worldline of a particle excitation $x$ winding around itself. Such information is encoded in the T-matrix.}   
\label{fig:ribbonT}  
\end{center}  
\end{figure}

For ${\cal C}$ a braided fusion category and $x\in {\cal C}$, the 2 morphisms

\begin{equation}  
x\ \xrightarrow{\text{id}_{_{x}}\otimes b_{_{x}}}\ x\otimes x\otimes x^*\ \xrightarrow{c_{_{x,x}}\otimes \text{id}_{_{x^*}}}\ x\otimes x\otimes x^*\xrightarrow{\text{id}_{_{x}}\otimes b_{_{x}}^{\dag}}\ x  
\end{equation}    

\begin{equation}  
x\ \xrightarrow{d_{_{x}}^{\dag}\otimes\text{id}_{_{x}}}\ x^*\otimes x\otimes x\ \xrightarrow{\text{id}_{_{x^*}}\otimes c_{_{x,x}}}\ x^*\otimes x\otimes x\xrightarrow{b_{_{x}}\otimes\text{id}_{_{x}}  }\ x  
\end{equation}  
are both equivalent to 

\begin{equation}  
\theta_{_{x}}:\ x\rightarrow x,  
\end{equation}  
defining the \emph{twist} or \emph{topological spin of x}. Importantly, $\theta_{_{x}}\neq\text{id}_{_{x}}$, since the worldline should be viewed as a ribbon, hence a line equipped with a framing. Both processes described in figure \ref{fig:ribbonT} are therefore equivalent to twisting the ribbon counterclockwise. The twisting rotates the topological defect by 360$^{\circ}$, reason why $\theta_{_{x}}$ is called the topological spin of x.

$\forall x\in {\cal C}$, since Hom$_{_{{\cal C}}}(x,x)\simeq\mathbb{C}$, 

\begin{equation}  
\theta_{_{x}}\equiv T_{_{x}}\cdot \text{id}_{_{x}}  
\ \ \ 
, \ \ \ 
T_{_{x}}\in\mathbb{C}, 
\end{equation}  
where the $T$-matrix of the unitary braided fusion category is defined as

\begin{equation}  
T\overset{def.}{=}
\left(T_{_{x}}\delta_{_{xy}}\right)_{_{x,y\in\text{Irr}({\cal C})}},
\label{eq:T}
\end{equation}  
encoding the information of self-statistics.

\subsection*{Quantum double category of a finite group }

$\forall$ finite group $G$, there is a unitary MTC ${\cal D}_{_{G}}$ called \emph{double quantum category of G}, \cite{Kong:2022cpy}, such that:  

\begin{enumerate}  

\item An object in ${\cal D}_{_{G}}$ is a vector space $V$ equipped with a $G$-action 

\begin{equation}   
\rho:\ G\ \longrightarrow\ GL(V),      
\end{equation}  
and a $G-grading$ $V\equiv \underset{g\in G}{\bigoplus} V_{_{g}}$ such that 

\begin{equation}  
\rho(g): V_{_{h}}\ \longrightarrow\ V_{_{ghg^{^{-1}}}}\ \ \ ,\ \ \ \forall g,h\ \in\ G.  
\end{equation}  

\item A morphism in ${\cal D}_{_{G}}$ is a $\mathbb{C}$-linear map that is, both, a morphism in Rep$(G)$ and a morphism in Vec$(G)$. 

\item  Given $g\in G$, we denote its conjugacy class as $[g]$, and its centraliser as $Z(g)$. If $\pi$ is an irrep $Z(g)$-representation, the induced representation Ind$_{_{Z(g)}}^{^{G}}(\pi)\overset{def.}{=} X_{_{(g,\pi)}}$ is a $G$-representation admitting a canonical $G/Z(g)\simeq[g]$-grading. $X_{_{(g,\pi)}}$ is a simple object in ${\cal D}_{_{G}}$.  

\item The isomorphism classes of objects in ${\cal D}_{_{G}}$ are labelled by pairs $([g], [\rho])$ where $[g]$ is the conjugacy class in $G$ and $[\rho]$ is an isomorphism class of irreps of $Z(g)$, where $Z[g]$ only depends on the conjugacy class of $g$.

\end{enumerate}

The properties of the $S$ and $T$ matrices defined in \eqref{eq:SM1} and \eqref{eq:T}, and their generalisations to the case of quantum double categories have been extensively described in previous works (see, for example, \cite{Kong:2022cpy}). For the purpose of the present work, it is important to stress that the $S$ and $T$ matrices are related to the generators of the $SL(2, \mathbb{Z})$ group 

\begin{equation} 
s\overset{def.}{=}\left(\begin{matrix}0\ \ -1\\
1\ \ \ \ 0\\ \end{matrix}\right)\ \ \ ,\ \ \ t\overset{def.}{=}\left(\begin{matrix}1\ \ \ 1\\
0\ \ \ 1\\ \end{matrix}\right),   
\end{equation}  
such that 

\begin{equation} 
(st)^{^3}\ \equiv\ s^{^2}\ \ \ ,\ \ \ s^{^4}\equiv 1.  
\end{equation}   
under a projective $SL(2, \mathbb{Z})$-representation

\begin{equation}  
s\ \mapsto\ \frac{S}{\ \sqrt{\text{dim}({\cal C})\ } } \ \ \ ,\ \ \ t\ \mapsto\ \ T.  
\label{eq:proj}   
\end{equation}  

The quantity ${\cal D}\overset{def.}{=}$dim$({\cal C})$ is the \emph{total quantum dimension} of the category ${\cal C}$. For the case in which ${\cal C}$ is made up of purely simple objects, the total quantum dimension of a given superselection sector simply reads  

\begin{equation}  
{\cal D}_{_{x}}\overset{def.}{=}\sqrt{\ \sum_{i}\ \text{dim}^{^2}(x_{_{i}})\ } =\sqrt {\ \sum_{i}\ S_{_{\mathbb{1}x_{i}}}^{^2}\ }.
\end{equation}

 \subsection{Example of a higher-categorical structure}    \label{sec:excatstr}

We now turn to describing the main example of MTC of interest to us. We will assume that the total category associated to a theory ${\mathfrak T}$ with gauge group $G$, ${\cal C}_{_{\mathfrak T}}$ is given by

\begin{equation}   
{\cal C}_{_{\mathfrak T}}\ \overset{def.}{=}\{{\cal C}_{_{point}}, {\cal C}_{_{flux}}, {\cal C}_{_{Hopf}}\ \},       
\label{eq:supersel}     
\end{equation}  
where each subcategory is a different superseclection sector of the theory, defined as follows:

\begin{enumerate}   

\item   ${\cal C}_{_{point}}\ \equiv \ \{1,a,...\}$, $a\equiv R\in\ (G)_{_{IRREP}}$, are point excitations, corresponding to the irreducible representations of the group $G$. The quantum dimension of each element of the category is $d_{_{a}}\ \overset{def.}{=}\ \text{dim}\ R$.

\item ${\cal C}_{_{flux}}\ \equiv\   \{1,\mu,...\}$ is the superselection sector of pure fluxes, and therefore corresponds to the conjugacy classes of the finite group $G$, $C\in\ (G)_{_{cj}}$. Their quantum dimensions are $d_{_{\mu}}\ \overset{def.}{=}\ \sqrt{|C|\ }$, with $C$ denoting the number of group elements in a given conjugacy class.

\item  ${\cal C}_{_{Hopf}}\ \equiv \ \{1,\eta,...\}$ is the collection of Hopf link excitations, 
with $\eta\overset{def.}{=}\ \left(C_{_{(g,h)}}, R\right), gh\equiv hg, C_{_{(g,h)}} \equiv\ \{(tgt^{-1}, tht^{-1}|t\in G\}, R\in\ E_{_{(g,h)}}\ \equiv\ \{t\in G\ |(g,h)\equiv \left(tgt^{-1}, tht^{-1}\right)\}$, and quantum dimensions 
\begin{equation}   
d_{_{\eta}}\ \overset{def.}{=}\ \frac{|G|}{\ |E_{_{(g,h)}}|\ }\ \text{dim}\ R   
\end{equation}   

\end{enumerate}

The 1-morphisms between these categories are

\begin{equation}  
\varphi\ :\ {\cal C}_{_{flux}}\ \hookrightarrow\ {\cal C}_{_{loop}}\ \subset\ {\cal C}_{_{Hopf}}  
\label{eq:varphi}
\end{equation}

\begin{equation}  
\phi\ :\ {\cal C}_{_{point}}\ \hookrightarrow\ {\cal C}_{_{loop}}\ \subset\ {\cal C}_{_{Hopf}}
\label{eq:phi}
\end{equation}

\begin{equation} 
v\ :\ {\cal C}_{_{Hopf}}\ \xrightarrow{\color{white}aaaa\color{black}}\ {\cal C}_{_{Hopf}}   \ \ \ , \ \ \ v(\eta)\ \equiv \eta^{^{\text{V}}}   
\label{eq:v}
\end{equation}

The aforementioned properties suggest the physical interpretation of the superselection sectors in \eqref{eq:supersel} as the electric, magnetic and dyonic charges characterising the spectrum of the theory $\mathfrak{T}$. As a result, the objects living in ${\cal C}_{_{\mathfrak T}}$ are defined by a 3-tuple $\chi\equiv([M],\lambda,\rho)$, corresponding to a flux-quasiparticle-charge composite.

As mentioned earlier on in the introduction, application of categories in physics began in the study of 2D CFTS and $D+1$ TQFTs. Their mathematical formulation was originally motivated by the idea of formalising the factorisation property of path-integrals as a certain monoidal functor defined on a cobordism category. Thanks to the pioneering work of Moore and Seiberg on\emph{modular tensor categories} in 2D RCFTs, Reshetikhin and Turaev reformulated their discovery in present-day categorical formulation, and used it to give the 2+1 D RT TQFT formulation, eventually leading to the study of topoplogical excitations (or anyons) in a 2D topological order. Anyons are quasiparticles, and as such can be thought of as being the physical counterpart of \eqref{eq:supersel}. Systems exhibiting anyon symmetries with gauge group $G$ in a given topological phase efficiently be related to Chern-Simons (CS) theory.

For such correspondence to hold, the CS theory needs to account for the topological information defined by the braiding statistics and quasiparticle fusion rules (cf. section \ref{sec:dfc}). For example, an abelian phase in (2+1)D can be characterised by a QFT with partition function defined as follows

\begin{equation}  
{\cal Z}[{\cal J}]\ \equiv\ \int\left[{\cal D}\alpha(\overset{\rightarrow}{r})\right]\ \exp\left(iS[{\cal J}] \right], 
\end{equation}  
involving an $N$-component set of $U(1)$ gauge fields $\alpha\equiv(\alpha_{_{1}},...,\alpha_{_{N}})$ with action

\begin{equation}  
S[{\cal J}] \equiv \frac{1}{4\pi}\ \int\left(K_{_{IJ}}\alpha^{^{I}}\wedge d\alpha^{^{J}}+\alpha^{^{I}}\ {\cal J}_{_{I}}\right).  
\label{eq:action}    
\end{equation}  

Quasiparticles, $\psi^{^{\overset{\rightarrow}{a}}}$, are sources for ${\cal J}_{_{1}}^{^{a_{_{1}}}},...,{\cal J}_{_{N}}^{^{a_{_{N}}}}$, labelled by $\overset{\rightarrow}{a} \equiv(a_{_{1}},...a_{_{N}})$ on a lattice $\Gamma^*\equiv\mathbb{Z}^{^N}$. At long distance, nearby quasiparticles can form a single entity, leading to the definition of a fusion structure. The $K$-matrix in \eqref{eq:action} dictates the braiding statistics of quasi-particles  

\begin{equation}
{\cal D}\ S_{_{ab}}\ \equiv\ e^{^{2\pi i a^{^T}K^{-1}b}}  
\ \ \ 
,   
\ \ \ 
{\cal D}\equiv\sqrt{|\text{det}(K)|\ }  \equiv\sqrt{{\cal A}\ }  
\end{equation} 
with ${\cal D}$ the total quantum dimension. The exchange statistics, instead, is defined by 

\begin{equation}  
\theta_{_{a}}\ \equiv\ e^{^{2\pi i a^{^T}K^{-1}b}},  
\end{equation} 
corresponding to the spin of the quasi particle. 
The topological phase generated by gauging the anyonic symmetry, is referred to as \emph{twisted liquid}. The latter are generalisations of (2+1)D discrete gauge theories. Their quasiparticles are compositions of fluxes and charges associated to the gauged anyonic symmetry, as well as a superselection sector of the original topological state. Their total quantum dimensions are related as follows 

\begin{equation}  
{\cal D}_{_{TL}}\ \equiv\ {\cal D}_{_{o}} \ |G|,  
\label{eq:DTL}   
\end{equation}  
with $|G|$ the order of the anyonic symmetry group $G$ being gauged. 

An important remark is of order, though. A CS description of a topological phase (cf. \eqref{eq:action}) is not unique. This follows from the fact that the $K$ matrix encodes the same fusion and braiding structure even after undergoing a basis transformation. The set of automorphisms, Aut$(K)$, preserving $K$ correspond an anyonic symmetry operation permuting the anyons with the same fusion properties and spin-statistics. As such, it classifies the global symmetries of the TQFT associated to the CS action, $S_{_{CS}}$. Once having modded-out the trivial relabellings, Inn$(K)$, we are left with the outer automorphisms 

\begin{equation}  
\text{Outer}(K)\ \overset{def.}{=}\ \frac{\text{Aut}(K)}{\text{Inner}(K)}, 
\end{equation}  
which will play a key role in the remainder of the present work.

These results about the total quantum dimension in gauged topological orders have been thoroughly investigated in recent years. In the next section, we shall see how the presence of intrinsic non-invertibles arising in the gauge theory lead to a different expression for the total quantum dimension, thereby signalling that the multiplicity of some superselection sectors can be greater that unity.

\section{Fiber functors from non-invertible symmetries} \label{sec:Freed-Moore-Teleman}   

The main motivation of the present section is that of studying the defects interpolating different 4D class ${\cal S}$ theories descending from dimensional reduction of 6D ${\cal N}$=(2,0) SCFTs. In particular, explaining how such defects emerge under gauging of multiple subalgebras, we propose the quantum dimension of the relative gaugeable subalgebra as the quantity needed to distinguish whether such interpolating defect is intrinsically or non-intrinsically non-invertible, thereby proposing the 4D counterpart of arguments proposed in \cite{Kaidi:2022cpf} for the 2D case.

The present section is therefore structured into 4 main parts:  

\begin{itemize} 

\item At first, we describe the so-called gauging-by-gauging procedure, \cite{TJF,Kong:2013aya,Yu:2021zmu}, and its adaptation to the Freed-Teleman, \cite{Freed:2012bs}, and Freed-Moore-Teleman construction, \cite{Freed:2022qnc,Freed:2022iao}, of relative field theories and topological symmetries of QFTs, respectively, with particular emphasis on the case of multiple simultaneous gauging of different subalgebras. 

\item We then briefly overview the procedure adopted by \cite{Bashmakov:2022uek} for realising absolute 4D SCFTs from ${\cal N}=(2,0)$ SCFTs, explicitly rephrasing it in terms of the multiple gauging-by-gauging procedure described in section \ref{sec:FMT}. 

\item When describing such interpolating defects in terms of gauging of homomorphisms, they naturally admit an interpretation in terms of fusion tensor categories interpolating between different 2-categories, to a total quantum dimension can be assigned. Section \ref{sec:2} closes with the relation between the issue in defining a fiber functor in presence of non-invertible defects interpolating between different absolute theories, which is current work in progress by the same author, \cite{VP}.

\item Making use of the gauging-by-gauging procedure explained in section \ref{sec:FMT}, we propose a criterion for distinguishing between the two types of non-invertible defects in terms of the total quantum dimension of the relative gauging algebra implementing the gauging. 

\end{itemize}

\subsection{Gauging-by-gauging in relative QFTs}  \label{sec:FMT}

\begin{figure}[ht!]  
\begin{center}
\includegraphics[scale=0.9]{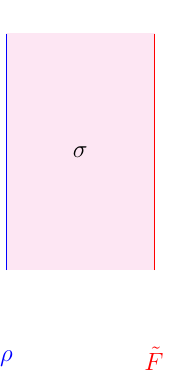} 
\ \ \ \ \ \ \ \ 
\includegraphics[scale=0.9]{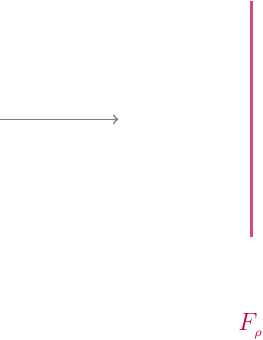}  
\caption{\small The Freed-Moore-Teleman setup, with $\tilde F$ denoting a relative QFT. Specifying the topological data $(\sigma,\rho)$, the resulting theory, $\tilde F_{_{\rho}}$ is absolute.}
\label{fig:FMT}    
\end{center}  
\end{figure}

In the formulation of \cite{Freed:2012bs}, a \emph{relative} field theory, $\tilde F$, requires additional topological data in order to be fully specified. Such data is encoded in a pair $(\sigma,\rho)$, referred to as \emph{quiche}. $\sigma$ is the \emph{symmetry topological field theory} (SymTFT), whereas $\rho$ geometrises the choice of boundary conditions for the fields defining the relative theory $\tilde F$. The overall system, depicted in figure \ref{fig:FMT}, gives rise to an \emph{absolute} QFT, $F_{_{\rho}}$.

In mathematical terms, the description outlined above can be formulated in terms of bordism in the following way. Fixing $N\in\mathbb{Z}^{^{\ge 0}}$, then a quiche is a pair $(\sigma, \rho)$ in which $\sigma:\text{Bord}_{_{N+1}}(F)\rightarrow{\cal C}$ is an $N+1$-dimensional TFT and $\rho$ is a right topological $\sigma$-module. The quiche is $N$-dimensional, hence it shares the same dimensionality as the theory on which it acts. Let $F$ be an $N$-dimensional field theory. A $(\sigma,\rho)$-\emph{module structure} on $F$ is a pair $(\tilde F, \theta)$, in which $\tilde F$ is a left $\sigma$-module and $\theta$ is an isomorphism 

\begin{equation}   
\theta\ : \rho\ \otimes_{_{\sigma}}\ \tilde F\xrightarrow{\ \ \simeq\ \ }\ F_{_{\rho}},  
\label{eq:LHS}  
\end{equation}   
of absolute $N$-dimensional theories, with \eqref{eq:LHS} defining the dimensional reduction leading to the absolute theory. $\sigma$ needs only be a \emph{once-categorified} $N$-dimensional theory, whereas $\rho$ and $\tilde F$ are relative theories.

In theories admitting a higher-categorical structure as \eqref{eq:supersel}, topological defects in different superselection sectors have different dimensionality. Consequently, the fusion rules are categorical, in the sense that they usually don't follow group-like composition laws. Because of this these defects are also called categorical or non-invertible. 

\begin{figure}[ht!]   
\begin{center}
\includegraphics[scale=0.7]{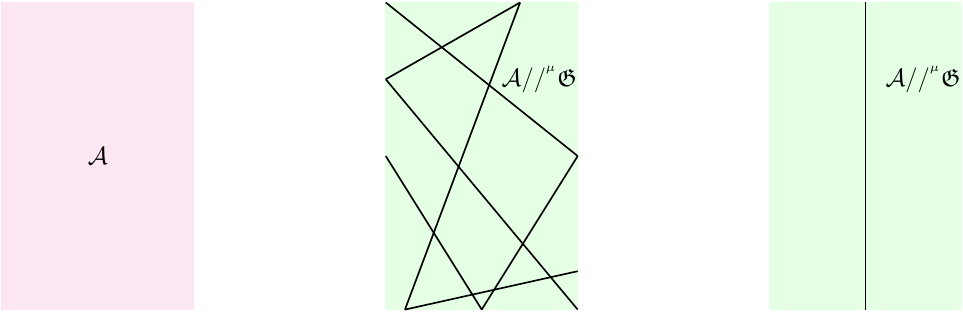}     
\caption{\small Gauging corresponds to gauging an algebra in a TFT. Idempotency ensures the resulting theory can be effectively thought of as featuring a unique defect, as shown on the RHS.}
\label{fig:codensationTJF}  
\end{center} 
\end{figure}

Prior to delving into the specific case of interest to us, namely relative QFTs, we first wish to provide some further explanation for what gauging a categorical structure actually means. In doing so, we refer to the work of many experts in the field, and, in particular \cite{TJF}. As explained in such reference, for any fusion n-category $\mathfrak{G}$, any fiber functor

\begin{equation} 
{\cal F}:\ \mathfrak{G}\ \rightarrow\ \text{nVec},  
\label{eq:functor}   
\end{equation}   
selects nVec as the image of a gauging algebra living in $\mathfrak{G}$, corresponding to a projection on the identity. The gauging process, can therefore be defined as a map

\begin{equation} 
\mu:\ \mathfrak{G}\ \rightarrow \ {\cal A}, 
\label{eq:mu}   
\end{equation}   
with ${\cal A}$ the algebra of invertible topological operators in $\mathfrak{G}$. Given \eqref{eq:mu}, the norm element

\begin{equation} 
N\overset{def}{=}\bigoplus_{g\in\mathfrak{G}}\ \mu(g).  
\label{eq:idemp}   
\end{equation}    
carries the structure of an n-categorical idempotent, also known as \emph{gauging algebra}, depicted in black in figure \ref{fig:codensationTJF}. The requirement for \eqref{eq:idemp} to be a higher-idempotent is needed to ensure the flooding doesn't depend on the specific features of the network being adopted to perform the gauging. The algebra of topological operators that are left are denoted by ${\cal A}//^{^{\mu}}\mathfrak{G}$. The equivalence of the second and third picture from the left in figure \ref{fig:codensationTJF} follows from $N$ being a higher-gauging algebra. As we shall see, this pattern emerges when gauging the 5D SymTFT of class ${\cal S}$ theories, where the objects of the 2-category in question will be Wilson surfaces charged under the 1-form symmetry being gauged.

\begin{figure}[ht!]   
\begin{center}
\includegraphics[scale=0.7]{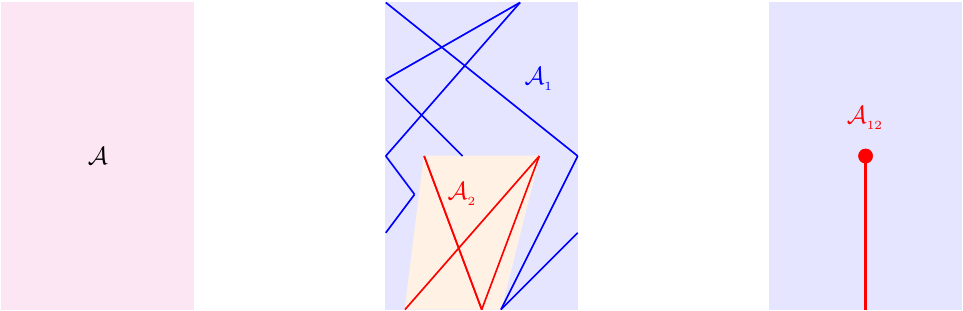}     
\caption{\small gauging two different subalgebras, ${\cal A}_{_1}, {\cal A}_{_2}\ \subset\ {\cal A}$, the resulting theory corresponds to one with a changed phase with a gauging defect resulting from a relative gaugeable algebra, ${\cal A}_{_{12}}$ ending in the bulk. The defect at the endpoint is nontrivial, and can therefore be thought of as a Hom$(\mathbb{1}_{_{{\cal C}}},{\cal A}_{_{12}})$.}  
\label{fig:2codensationsTJF}  
\end{center} 
\end{figure}

Unlike \ref{fig:codensationTJF}, \ref{fig:2codensationsTJF} does not admit a straightforward expression as \eqref{eq:functor}.

\begin{figure}[ht!]   
\begin{center}
\includegraphics[scale=0.9]{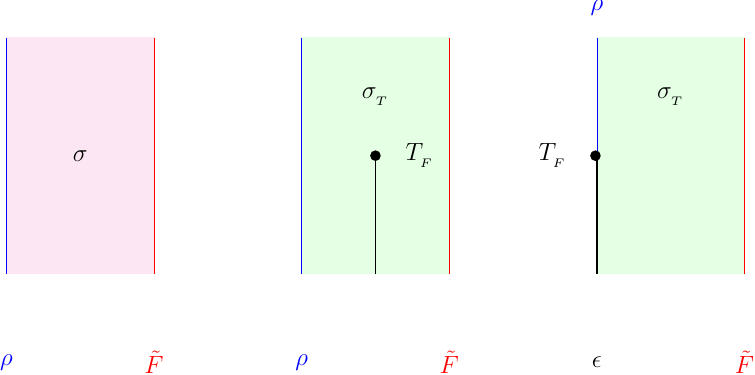}  \ \ \ \ \ \ \ \ \ \ \ \ \ \ \ \ \ \ 
\includegraphics[scale=1]{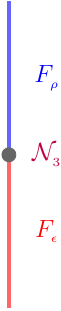} 
\caption{\small Adaptation of Freed-Moore-Teleman to the case involving twisted gauging defects. The figure on the far right corresponds to the case of interest to us, namely a configuration involving two different absolute 4D gauge theories separated by a defect. As we shall see, such defect is intrinsically non-invertible, corresponding to the presence of a relative ungauged subalgebra, dressing ${\cal N}_{_3}$ in a nontrivial way. In higher-categorical terms, it corresponds to a fusion tensor category implementing the morphisms between the operators charged under the gauged symmetry.}
\label{fig:FrredMooreTeleman extended}  
\end{center} 
\end{figure}

\subsection{Absolute 4D theories from 6D \texorpdfstring{${\cal N}=(2,0)$}{} SCFTs}    \label{sec:4dfrom6d}

The procedure described in section \ref{sec:FMT} cannot be applied to the case of maximally supersymmetric 6D theories. The reason being that, while a bulk SymTFT can be defined, thanks to the Milnor-Moore theorem, specifying any topological boundary conditions would inevitably imply supersymmetry breaking, \cite{Freed:2012bs}. Nevertheless, such 6D theories provide the natural realm from which one can realise lower-dimensional absolute theories, as well as constituting an extremely rich setting where to explore and put to practice mathematical techniques provided by representation theory, \cite{DBZ}.

The main features of 6D ${\cal N}$=(2,0) SCFTs are that:  

\begin{itemize}  

\item They admit no Lagrangian formulation, \cite
{Freed:2012bs, Witten:2009at}.  

\item The theory depends on the Lie algebra, $\mathfrak{g}$, rather than the Lie group, $G$. 

\item Ordinary\footnote{By this we mean that the lower-dimensional theories depend on the specific choice of the Lie group.} 5D and 4D QFTs can be obtained by dimensional reduction on a Riemann surface, $\Sigma_{_{g,n}}$.  

\item It is \emph{relative} w.r.t. an \emph{extended} TFT in 1-dimension higher,  \cite
{Freed:2012bs}. 

 \end{itemize}

Assuming the 6D ${\cal N}=(2,0)$ SCFT is defined on $Y=X\times \Sigma_{_{g}}$, with $X$ a compact 4D spacetime, torsion-free, and with trivial 1$^{^{st}}$ cohomology, $H^{^{1}}(X,\mathbb{Z}_{_{N}})$, the 6D self-dual fluxes belong to\footnote{$\mathbb{Z}_{_{N}}$ denotes the \emph{defect group} of the 6D theory.}

\begin{equation}  
H^{^{3}}(Y,\mathbb{Z}_{_{N}})\ \simeq\ H^{^{1}}(\Sigma_{_{g}},\mathbb{Z}_{_{N}})\ \otimes\ H^{^{2}}(X,\mathbb{Z}_{_{N}}).   
\end{equation}     

As far as the 6D theory is concerned, this is the only information at our disposal, given that the theory is non-Lagrangian, \cite{Freed:2012bs, Witten:2009at}. However, for the 4D theory to be fully defined, additional data is needed. In particular, one needs to identify a \emph{maximal isotropic sublattice} $L$ of $H^{^{1}}(\Sigma_{_{g}},\mathbb{Z}_{_{N}})$ w.r.t. the canonical pairing induced by the intersection pairing on $\Sigma_{_{g}}$, thereby leading to 

\begin{equation}  
{\cal L}\ \simeq\ L\ \otimes\ H^{^{2}}(X,\mathbb{Z}_{_{N}})   
\label{eq:maxissub}    
\end{equation} 
as the maximal isotropic lattice of $H^{^{3}}(Y,\mathbb{Z}_{_{N}})$. \eqref{eq:maxissub} enables to distinguish between different global forms of class ${\cal S}$ theories. Having fixed ${\cal L}$, the fluxes in its complement

\begin{equation} 
{\cal L}^{\perp}\  \overset{def.}{=}\ H_{_{3}}(X_{_{6}},\mathbb{Z}_{_{N}})/{\cal L}  
\end{equation} 
parametrise the possible partition functions for the 4D theory with inequivalent 1-form symmetry backgrounds along $X$. 

The 6D ${\cal N}$=(2,0) theory, assigns to the 6D manifold $Y$ a partition vector rather than a partition function, with the former belonging to the Hilbert space constituting a representation of a Heisenberg algebra of non-commuting discrete 3-form fluxes in $\mathbb{Z}_{_{N}}$.  

For fixed ${\cal L}$, the partition vector takes the form

\begin{equation}  
|{\cal Z}(Y)>\ \equiv\ \sum_{v\ \in\ {\cal L}^{^{\perp}}}\ {\cal Z}_{_{v}}(Y)\ |{\cal L}; v>,   
\label{eq:chofb}   
\end{equation}    
with ${\cal Z}_{_{v}}(Y)$ denoting the 6D conformal blocks, which also encode the \emph{global structures} of the 4D theory. Under a change of basis, \eqref{eq:chofb} turns into   

\begin{equation}  
|{\cal Z}(Y)>\ \equiv\ \sum_{v^{\prime}\ \in\ {\cal L}^{^{\prime\perp}}}\ {\cal Z}_{_{v}}(Y)\ \sum_{v\ \in\ {\cal L}^{^{\perp}}}\ R_{_{v}}^{^{v^{\prime}}}\ |{\cal L}; v>,  
\label{eq:chofb1}   
\end{equation} 
leading to the following relation between conformal blocks

\begin{equation}  
{\cal Z}_{_{v}}(Y)\ \equiv\ \sum_{v^{\prime}\ \in\ {\cal L}^{^{\prime\perp}}}\ R_{_{v}}^{^{v^{\prime}}}\ {\cal Z}_{_{v}}(Y).     
\label{eq:chofb2}   
\end{equation}

\section*{The 7D SymTFT}

The Milnor-Moore Theorem states that, the universal enveloping algebra of a Lie algebra is a Hopf algebra whose primitive elements are the elements of the original Lie algebra. A Hopf algebra on an associative algebra turns the category of modules into a monoidal category equipped with a fiber functor. This is a statement of Tannaka duality. 

Putting this together for the case of 6D ${\cal N}=(2,0)$ SCFTs, the bulk SymTFT is a pointed braided tensor category where the braiding is in bijection with 3-cocycles.

Following the arguments outlined in section \ref{sec:excatstr}, one can take the 7D TQFT in the bulk of the relative 6D ${\cal N}=$(2,0) theory, to be a CS theory with action

\begin{equation}  
S_{_{7D}}=\frac{N}{4\pi}\ \int_{_{W_{_{7}}}}\ dc\ \wedge\ c\ \ \ ,\ \ \ c\in H^{^{3}}(W_{_{7}},U(1)), \label{eq:7D}   
\end{equation}   
and Wilson surfaces  

\begin{equation}  
\Phi_{_{q}}({\cal M}_{_{3}})\overset{def.}{=}\ e^{^{iq\oint_{_{{\cal M}_{_{3}}}} c}}\ \ \ ,\ \ \ q\in\mathbb{Z}_{_{N}}\ \ \ ,\ \ \ {\cal M}_{_{3}}\ \in H^{^{3}}(W_{_{7}},U(1)). 
\end{equation}   

Taking two such Wilson surfaces $\Phi_{_{q}}({\cal M}_{_{3}}), \Phi_{_{q^{\prime}}}({\cal M}_{_{3}}^{\prime})$, with ${\cal M}_{_{3}}, {\cal M}_{_{3}}^{\prime}$ forming a Hopf link in $W_{_{7}}$, and inserting them in the path integral, amounts to changing the action \eqref{eq:7D} by adding the holonomy terms associated to the operator insertions

\begin{equation}  
\begin{aligned}
S_{_{7D}}&=\frac{N}{4\pi}\ \int_{_{W_{_{7}}}}\ dc\ \wedge\ c  + q\int_{_{{\cal M}_{_{3}}}} c\ +q^{\prime} \int_{_{{\cal M}_{_{3}}^{\prime}}} c\\  
&=\frac{N}{4\pi}\ \int_{_{W_{_{7}}}}\ dc\ \wedge\ c +\ \int_{_{W_{_{7}}}}\ \left(q\omega_{_{{\cal M}_{_{3}}}}+q^{\prime}\omega_{_{{\cal M}_{_{3}}^{\prime}}}\right)\ \wedge\ c,
\label{eq:CS7D}    
\end{aligned}
\end{equation} 
with $\omega_{_{{\cal M}_{_{3}}}}, \omega_{_{{\cal M}_{_{3}}^{\prime}}}$ denoting the Poincar\'e duals of ${\cal M}_{_{3}}, {\cal M}_{_{3}}^{\prime}$, respectively. Integrating-out $c$, and imposing 

\begin{equation}  
dc\equiv-\frac{2\pi}{N}\left(q\omega_{_{{\cal M}_{_{3}}}}+q^{\prime}\omega_{_{{\cal M}_{_{3}}^{\prime}}}\right), 
\end{equation} 
defining $V_{_{4}}$ such that 

\begin{equation} 
\partial V_{_{4}}\equiv\left(q\omega_{_{{\cal M}_{_{3}}}}+q^{\prime}\omega_{_{{\cal M}_{_{3}}^{\prime}}}\right),  
\end{equation} 
implies

\begin{equation}   
c\equiv -\frac{2\pi}{N}\ \text{PD}(V_{_{4}}).  
\end{equation}

From this, \eqref{eq:CS7D} reduces to 

\begin{equation}  
\begin{aligned}
S_{_{7D}} 
&=\frac{2\pi}{N}\ q q^{\prime}\ \text{link}\left({\cal M}_{_{3}}, {\cal M}_{_{3}}^{\prime}\right),
\label{eq:linkact7}  
\end{aligned}
\end{equation}
where the linking is the Hopf link occurring in $W_{_{7}}$. It features in \eqref{eq:linkact7} due to the fact that $V_{_{4}}$ is a Seifert surface for the combination $q{\cal M}_{_{3}}+q^{\prime}{\cal M}_{_{3}}^{\prime}$, meaning that, every time ${\cal M}_{_{3}}$ pierces $V_{_{4}}$, it links ${\cal M}_{_{3}}^{\prime}$ once. It thereby follows that, the Hopf link of Wilson surfaces supported on ${\cal M}_{_{3}}$ and ${\cal M}_{_{3}}^{\prime}$ in $W_{_{7}}$ can be thought of as an operator insertion in the 7D path integral, 

\begin{equation}  
<\Phi_{_{q}}({\cal M}_{_{3}})\Phi_{_{q^{\prime}}}({\cal M}_{_{3}}^{\prime})...>\equiv e^{^{\frac{2\pi i}{N}q q^{\prime}\text{link}({\cal M}_{_{3}},{\cal M}_{_{3}}^{\prime})}}\ <...>,  
\end{equation}  
where the phase on the RHS being the analog of the $S$-matrix element $S_{_{ab}}$ between anyons of charges $a,b$ in $U(1)_{_{k}}$ CS theory in 3D, namely $S_{_{ab}}\equiv e^{^{2\pi iab/k}}$.

For a given $X_{_{6}}\equiv \Sigma_{_{g,n}}\times X_{_{4}}$, there is a particular class of 7-manifolds $W_{_{7}}$ obtained by taking $W_{_{7}}\equiv V_{_{g,n}}\times X_{_{4}}$, with $V_{_{g,n}}$ a 3-manifold with $\partial V_{_{g,n}}\equiv\Sigma_{_{g,n}}$. 

For any Riemann surface $\Sigma_{_{g,0}}$, there are many inequivelent 3-manifolds with $\Sigma_{_{g,0}}$ as its boundary. One such example are handlebodies. To construct a handlebody, choose a set $g$ of \emph{meridians}, $\{\mu_{_{i}}\}, i\equiv1,...g$, i.e. a set of generators of $H_{_{1}}(\Sigma_{_{g,0}}, \mathbb{Z})$ which become trivial as elements of $H_{_{1}}(V_{_{g,0}}, \mathbb{Z})$.  The remaining $g$ generators of $H_{_{1}}(\Sigma_{_{g,0}}, \mathbb{Z})$ lift to generators of $H_{_{1}}(V_{_{g,0}}, \mathbb{Z})$, and are referred to as \emph{longitudes}, $\{\lambda_{_{i}}\}, i\equiv1,...g$. Not every choice of $g$ generators gives rise to a legitimate set of meridians (i.e. vanishing cycles). For a specific choice of meridians to be legitimate, it must correspond to a maximal isotropic sublattice   

\begin{equation}  
L\ \in\ H_{_{1}}(\Sigma_{_{g,0}},\mathbb{Z}). 
\end{equation}  

The handlebody specified by the choice of meridians $L$ will be denoted by $V_{_{g,0}}^{^{L}}$. Given $V_{_{g,0}}^{^{L}}$, there are multiple choices of longitudes, differing by shifts in meridians   

\begin{equation}  
\lambda_{_{i}}^{\prime}\equiv\lambda_{_{i}}+\sum_{j=1}^{3}k_{_{ij}}\mu_{_{j}}\ \ \ ,\ \ k_{_{ij}}\in\mathbb{Z},
\end{equation}   
satisfying the following constraint 

\begin{equation}
<\mu_{_{i}},\lambda_{_{j}}>\equiv-<\lambda_{_{j}},\mu_{_{i}}>\equiv\delta_{_{ij}} . 
\end{equation}   

7D CS theory on a handlebody can be formulated by starting with a definition of the fields which reads 

\begin{equation}   
b_{_{i}}\overset{def.}{=}\oint_{_{\mu_{_{i}}}} c\ \ \ ,\ \ \ \hat b_{_{i}}\overset{def.}{=}\oint_{_{\lambda_{_{i}}}} c\ \ \ ,\ \ \ i=1,...,g.  
\end{equation}  

The meridians are naively contractible by definition. However, this does not necessarily imply the vanishing of $b_{_{i}}$, since there might be Wilson surfaces supported on $\lambda_{_{i}}$ and $\mu_{_{i}}$ that are nontrivially linking, effectively making $\mu_{_{i}}$ non-contractible. 

The Wilson lines wrapping meridians and longitudes can be expressed as follows  

\begin{equation}   
\Phi_{_{i}}({\cal M}_{_{2}})\ \overset{def.}{=} \Phi_{_{i}}({\cal M}_{_{2}}\times\mu_{_{i}})\equiv e^{^{\frac{2\pi i}{N}\oint_{_{{\cal M}_{_{2}}}}b_{_{i}}}}  
\ \ \ ,\ \ \ \hat \Phi_{_{i}}({\cal M}_{_{2}})\ \overset{def.}{=} \hat \Phi_{_{i}}({\cal M}_{_{2}}\times\lambda_{_{i}})\equiv e^{^{\frac{2\pi i}{N}\oint_{_{{\cal M}_{_{2}}}}\hat b_{_{i}}}},
\end{equation}   
which might be nontrivially linking in the 5D theory resulting from dimensional reduction of the 7D TFT on $\Sigma_{_{g,0}}$. Compactifying the theory on $\Sigma_{_{g,0}}$ in presence of a series of Wilson 3-surfaces on $\lambda_{_{j}}\times {\cal M}_{_{2,j}}^{\prime}$, the linking equation becomes an operator equation in 4D 

\begin{equation}  
\Phi_{_{i}}({\cal M}_{_{2}})\ \equiv\ e^{^{\frac{2\pi i}{N}\sum_{j=1}^{g}<\mu_{_{i}},\lambda_{_{i}}>\oint_{_{{\cal M}_{_{2}}}}B_{_{j}}}},  
\label{eq:extraphase}
\end{equation}  
where $B_{_{j}}\in H^{^{2}}(X_{_{4}},\mathbb{Z}_{_{N}})$ denotes the Poincar\'e dual of ${\cal M}_{_{2,j}}^{\prime}$. Hence, in any 4D calculation involving $\Phi_{_{i}}$, the latter can be moved to the right of $\hat\Phi_{_{j}}$, and subsequently be shrunk to a point by introducing the phase \eqref{eq:extraphase}. Hence, the insertion of longitudinal Wilson lines allows us to make $b_{_{i}}\neq 0$, while still keeping it constant, thereby defining a background field.

At this point, the Sym TFT can be obtained by splitting the handlebody in the following way:

\begin{equation}  
V_{_{g,0}}\ \overset{def.}\ \begin{cases} V_{_{g,0}}^{^{L,in}}\ \ \ \text{for}\ \ \ y\ge y_{_{*}}\\ 
\\
V_{_{g,0}}^{^{ext}}\overset{def}{=}\Sigma_{_{g,0}}\times[0,y_{_{*}}]\ \ \ \text{for}\ \ \ 0\le y\le y_{_{*}}\\ 
\end{cases}.  
\label{eq:splitting}  
\end{equation}  

\begin{figure}[ht!]   
\begin{center}  
\includegraphics[scale=0.7]{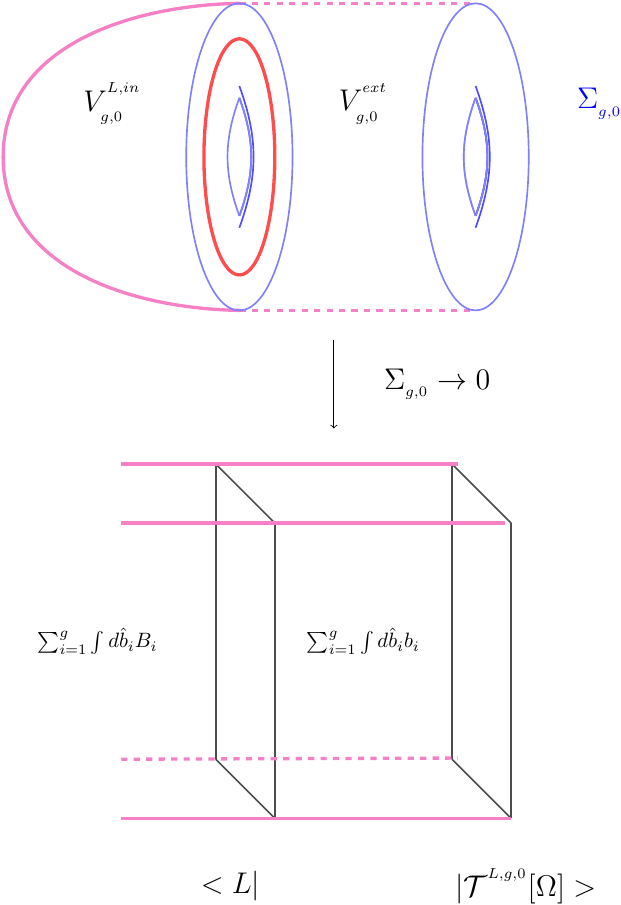}   
\caption{\small In presence of the splitting \eqref{eq:splitting}, the state $<L|$ separates the anomaly TFT from the Sym TFT. The picture above is valid for $\forall g$, but has been drawn with $g=1$ for simplicity.}    
\label{fig:shrsigma}  
\end{center}    
\end{figure}

Importantly, it features two boundaries (cf. figure \ref{fig:shrsigma}): 

\begin{itemize} 

\item One, at $y=0$, is where the 6D ${\cal N}=$(2,0) theory lives. 

\item The other, at $y=y_{_{*}}$, is where the longitudinal Wilson surfaces fixing the background fields are placed. 

\end{itemize}

  \subsection{Intrinsic non-invertible symmetries and fiber functors} \label{sec:2}

On the handlebody, one can perform two different operations:  

\begin{enumerate}  

\item A \emph{modular transformation} of $\Sigma_{_{g,0}}$, acting on the entire handlebody and generically changing the period matrix of $\Sigma_{_{g,0}}$ at each cross section. The theory in the bulk of the handlebody is topological, implying only $\Sigma_{_{g,0}}$ is affected by such transformations, which is where the 6D ${\cal N}=$(2,0) theory lives. 

\item Excising an \emph{inner} handlebody, in the setup where the splitting \eqref{eq:splitting} takes place. The operation of interest consists in gluing together the 2 parts $V_{_{g,0}}^{^{L,in}}$ and $V_{_{g,0}}^{^{ext}}$ with a nontrivial element of the modular group $Sp(2g, \mathbb{Z}_{_{N}})$, corresponding to a surgery operation on the 3-manifolds. 

\end{enumerate}    

Given that the latter takes place in the interior of the handlebody, it doesn't affect $\Sigma_{_{g,0}}$ where the 6D theory lives, hence, the period matrix remains unchanged in the region the theory is sensitive to. However, this surgery operation changes the global form of $L$, since it changes which are the contractible cycles in the handlebody. Hence, surgery is implemented by an element $F$ of the modular group which only affects the inner boundary of the handlebody, thereby changing the global form. This operation may in turn be combined with a modular transformation acting on the entire handlebody, reverting the interior of the handlebody back to its original form, but now changing the period matrix of the Riemann surface. 

Choosing a period matrix that remains invariant under the action of $F$, the geometry with the combined surgery plus modular transformation has the same boundaries as the original geometry. Importantly, though, the full geometry is no longer the same, due to the internal twist.

Under dimensional reduction on $\Sigma_{_{g,0}}$, \eqref{eq:CS7D} becomes    

\begin{equation}  
S_{_{5D}}\equiv\frac{2\pi}{N}\sum_{i=1}^{g}\int_{_{X_{_{4}\times[0,y_{_{*}}]}}}b_{_{i}}\ \cup\ \delta\hat b_{_{i}}. 
\label{eq:newBFnew}   
\end{equation}

The fact that $b_{_{i}}, \hat b_{_{i}}$ are dynamical, implies $\lambda_{_{i}}, \mu_{_{i}}$ are both nontrivial in $V_{_{g,0}}^{^{ext}}$. This follows from the fact that $L$ specifies which directions of the Riemann surface become trivial in the handlebody, namely the meridians, and which of the dynamical fields of the $BF$-theory become background. To each $L$ corresponds a different choice of Dirichlet boundary conditions for different sets of fields in the $BF$-theory. As explained in \cite{Bashmakov:2022uek}, the 7D CS theory on a Riemann surface can only capture the full SymTFT iff all the non-invertible defects are non-intrinsic.  

The 5D Wilson lines are defined from the 7D Wilson surfaces as follows

\begin{equation}  
\Phi_{\overset{\rightarrow}{n}}\left({\cal M}_{_{2}}\right)\ \overset{def.}{=}\  \Phi\left({\cal M}_{_{2}}\times\gamma_{_{\overset{\rightarrow}{n}}}\right),
\label{eq:phim2}   
\end{equation}
with 1-cycles 

\begin{equation}  
\gamma_{_{\overset{\rightarrow}{n}}}  \ \overset{def.}{=}\ \sum_{j=1}^{g} e_{_{j}}\lambda_{_{j}}+\sum_{j=1}^{g}\ m_{_{j}}\mu_{_{j}}\ \ \ ,\ \ \ \overset{\rightarrow}{n}\ \overset{def.}{=}\ \left(e_{_{1}},..., e_{_{g}};m_{_{1}},...,m_{_{g}}\right).
\end{equation}

Explicitly, \eqref{eq:phim2} can be expressed as follows  

\begin{equation}  
\Phi_{\overset{\rightarrow}{n}}\left({\cal M}_{_{2}}\right)\ \equiv\  e^{^{\frac{2\pi i}{N}\ \frac{1}{2}e\cdot m({\cal M}_{2},{\cal M}_{2})}}\ \prod_{i=1}^{g}\Phi^{^{e_{_{i}}}}\left({\cal M}_{_{2}}\times\gamma_{_{\overset{\rightarrow}{n}}}\right) \ \ \prod_{i=1}^{g}\Phi^{^{m_{_{i}}}}\left({\cal M}_{_{2}}\times\gamma_{_{\overset{\rightarrow}{n}}}\right),  
\label{eq:phim21}   
\end{equation}
with each electric and magnetic component, in turn descending from a 3D Wilson surface in 7D associated to a cycle $\gamma_{_{\overset{\rightarrow}{n}}}$. 

Surgery defects in 7D correspond to loci across which $F$ acts on $\gamma_{_{\overset{\rightarrow}{n}}}$ as follows 

\begin{equation}   
F\left(\gamma_{_{\overset{\rightarrow}{n}}}\right)\ \equiv\ \gamma_{F\overset{\rightarrow}{n}}.  
\end{equation}

In 5D, they reduce to gauging defects implementing morphisms in a higher-categorical structure

\begin{equation}  
{\cal C}_{_{F}}(X_{_{4}})\ :\ \Phi_{\overset{\rightarrow}{n}}\left({\cal M}_{_{2}}\right)\ \mapsto\ \Phi_{F\overset{\rightarrow}{n}}\left({\cal M}_{_{2}}\right)\ \ \ \ ,\ \ \ {\cal M}_{_{2}}\ \subset\ X_{_{4}}.  
\label{eq:morphism}  
\end{equation}

\begin{figure}[ht!]   
\begin{center}
\includegraphics[scale=0.8]{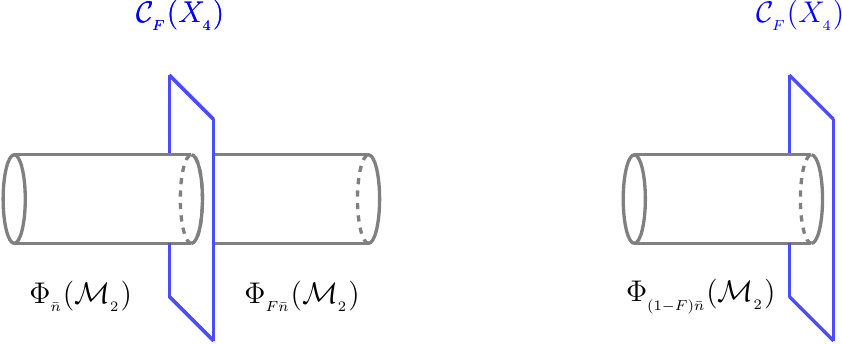}  
\caption{\small The picture on the LHS shows the identification of a morphism between objects as a defect in a higher-categorical structure. On the RHS, the interface turns into a boundary defect once having performed the folding trick.}
\label{fig:jphi}  
\end{center} 
\end{figure}

\begin{figure}[ht!]    
\begin{center}   
\includegraphics[scale=0.8]{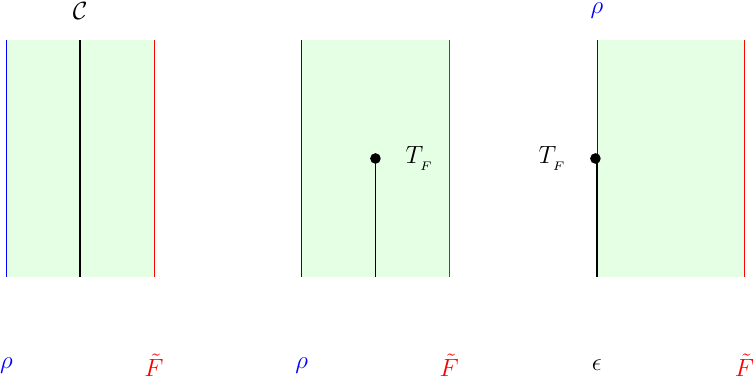}  
\caption{\small Gauging the defects living in the SymTFT upon implementing an outer automorphisms, changes the boundary conditions of the resulting absolute theory. In presence of a unique gauging, the resulting configuration is depicted on the LHS. On the other hand, upon gauging two different subalgebras of the original bulk theory, leads to a bulk gauging defect effectively ending in the bulk, as shown in the figure in the middle. This is equivalent to realising a configuration comprising two different boundary conditions separated by a topological defect (as shown on the RHS).}   
\label{fig:newidentity}   
\end{center}   
\end{figure}

As suggested by \eqref{eq:morphism}, $F$ acts on the labellings of the objects living in the theory. These objects are dyonic, with nontrivial linking. An explicit expression for the morphism \eqref{eq:morphism} as a defect follows by performing the folding trick on the surgery point in the 5D configuration depicted in figure \ref{fig:jphi}

\begin{equation}  
{\cal C}(X_{_{4}}) \ =\ \frac{\ |H^{^{0}}(X_{_{4}},\mathbb{Z}_{_{p}})|\ }{|H^{^{1}}(X_{_{4}},\mathbb{Z}_{_{p}})|}\ \sum_{_{{\cal M}_{_{2}} \in H_{_{2}}\left(X_{_{4}}, \mathbb{Z}_{_{p}}^{^{2g}}\right)}}\ \exp\left(\frac{2\pi i}{p}\ <F{\cal M}_{_{2}},{\cal M}_{_{2}}>\right) \Phi \left((\mathbb{1}-F){\cal M}_{_{2}} \right),    
\label{eq:conddef}  
\end{equation} 
implying that, $\forall \ F\in\ Sp(2g, \mathbb{Z}_{_{p}})$, there is a codimension-1 defect implementing the symmetry associated to $F$ in the bulk 5D TQFT. Such defect is invertible, and therefore its fusion rules simply read

\begin{equation}   
\boxed{\ \ \ {\cal C}(X_{_{4}})\ \times\ \bar {\cal C}(X_{_{4}})\ \equiv\ {\cal A}\ \equiv\ \mathbb{1}\color{white}\bigg]\ \ }. 
\label{eq:identity}
\end{equation}
where ${\cal A}$ corresponds to  the gauged algebra implementing the gauging. Notice that the defect \eqref{eq:conddef} is equivalent to the one arising from gauging by algebraic gauging in a topological order, as explained in section \ref{sec:FMT} and depicted in figure \ref{fig:codensationTJF}. The crucial meaning of equation \eqref{eq:identity} is the fact that, gauging-by-gauging is effectively acting as a redefinition of the identity from the mother theory to the gauged one. In particular, the last equality claims that the gauged algebra in the mother theory is projected to the identity of the gauged theory. Pictorially, this corresponds to the configuration on the LHS of \ref{fig:newidentity}.

Gauging defects associated to $F$, can in turn give rise to $|F|$-ality defects in the boundary 4D gauge theory. Such defects can be constructed from \eqref{eq:conddef} by admitting Dirichlet boundary conditions for the gauged defects, allowing them to terminate in the bulk, \cite{Kaidi:2022cpf, Teo:2015xla, Barkeshli:2014cna}. Practically, this corresponds to a non-genuine 3D-manifold attached to a 4D manifold, which, combined together, give rise to a \emph{twist defect}, $T_{_{F}}({\cal M}_{_{3}}, {\cal M}_{_{4}})$, whose expression is obtained from \eqref{eq:conddef} once rewritten in terms of relative cohomology. From the perspective of the gauging-by-gauging procedure, this amounts to considering simulatneous gauging of two different subalgebras within the gaugeable subalgebra of the mother theory. In terms of its SymTFT characterisation, this would correspond to the gauging defect effectively ending in the bulk SymTFT as shown in the middle of figure \ref{fig:newidentity}. Equivalently, this is equivalent to describing a relative field theory with two boundary conditions interpolated by a topological defect.

\begin{figure}[ht!]    
\begin{center}   
\includegraphics[scale=0.8]{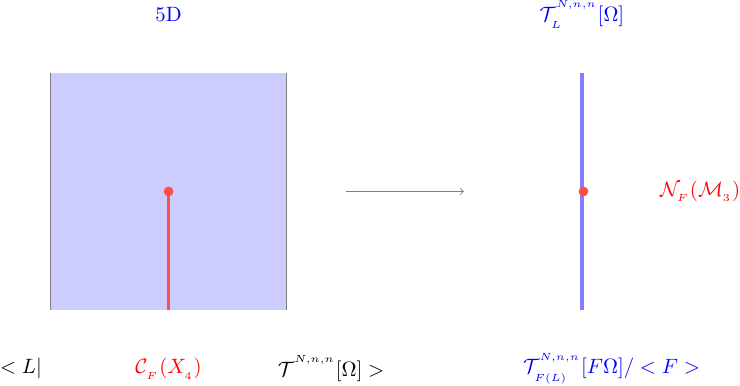}  
\caption{\small The gauging defect ${\cal C}_{_{F}}(X_{_{4}})$ implements the outer automorphism in the 5D TQFT. Applying twisted BCs, the gauging defect can effectively terminate in the 5D bulk, and is therefore topological. At this point, the width of the 5D theory can be taken to vanish, resulting in the configuration on the RHS, with the two phases of the 4D theory separated by a topological non-invertible defect, ${\cal N}_{_{F}}({\cal M}_{_{3}})$.}   
\label{fig:gaugingdefect}   
\end{center}   
\end{figure}

The gauging defect can be re-expressed in terms of the quantum dimensions of the underlying anyon description by noticing that the action of $F$ is equivalent to that of the outer automorphism keeping track of the anyonic symmetry being gauged\footnote{As explained in section \ref{sec:excatstr} and \ref{sec:FMT}.}. This follows from the fact that the elements of the sum appearing in the gauging defect, \eqref{eq:conddef}, can be broken down into two main parts, namely     

\begin{equation}  
\exp\left(\frac{2\pi i}{p}\ <F{\cal M}_{_{2}},{\cal M}_{_{2}}>\right) \Phi \left((\mathbb{1}-F){\cal M}_{_{2}} \right)\ \equiv\ S_{_{\overset{\rightarrow}{a}\ F\overset{\rightarrow}{a}}}\ {\cal G}_{_{F\overset{\rightarrow}{a}}},  
\end{equation}  
with $S_{_{\overset{\rightarrow}{a},\ F\overset{\rightarrow}{a}}}$ and ${\cal G}_{_{F\overset{\rightarrow}{a}}}$ denoting the $S$-matrix entries for the associated anyonic charges, and the exceptional fibration operator\footnote{A deeper explanation of this is the key topic of an upcoming work \cite{VP}.}, respectively. Extending these arguments, the sum over the 2-manifolds ${\cal M}_{_{2}}$ can be replaced by a sum over the anyonic charges compatible with the symmetries of the given theory. This way, \eqref{eq:conddef} can be rewritten as follows

\begin{equation}  
T_{_{F}}({\cal M}_{_{3}}, {\cal M}_{_{4}}) \ \overset{def.}{=}\ \frac{\ |H^{^{0}}(X_{_{4}},\mathbb{Z}_{_{p}})|^{^{2g}}\ }{|H^{^{1}}(X_{_{4}},\mathbb{Z}_{_{p}})|^{^{2g}}}\ \sum_{_{\overset{\rightarrow}{a}}} \ S_{_{\overset{\rightarrow}{a},\ F\overset{\rightarrow}{a}}}\ {\cal G}_{_{F\overset{\rightarrow}{a}}},
\label{eq:conddef1}  
\end{equation} 
where $\overset{\rightarrow}{a}\ \overset{def.}{=}\ \left([a], \rho(a)\right)$, with $[a], \rho(a)$ denoting the conjugacy class and the representation of the corresponding anyon charges, respectively. 
Given that the defect is now topological, it can be moved to the 4D boundary, defining an interface for the 4D gauge theory

\begin{equation}  
{\cal N}({\cal M}_{_{3}})\ \overset{def.}{=}\ \underset{y\rightarrow0}{\lim}\ T_{_{F}}({\cal M}_{_{3}}, {\cal M}_{_{4}}),
\end{equation}   
obeying non-invertible fusion rules. From the 4D perspective of the absolute theory on the RHS of figure \ref{fig:gaugingdefect}, ${\cal N}({\cal M}_{_{3}})$ is \emph{intrinsically}-non-invertible, namely, it cannot be rendered invertible by acting with $SL(2,\mathbb{Z})$-transformations. This is equivalent to stating that the action of $F$ cannot be reversed simply by acting with a modular transformation. Non-invertibility implies its fusion rules now read 

\begin{equation}  
\boxed{\ \ \ {\cal N}({\cal M}_{_{3}})\ \times \bar{\cal N}({\cal M}_{_{3}})\ =\ {\cal A}_{_{X_{_{4}}}}\color{white}\bigg]\ \ },   
\label{eq:NNA}   
\end{equation}
with ${\cal A}_{_{X_{_{4}}}}$ denoting the relative gaugeable algebra in the absolute theory. 

Crucially, the RHS of equation \eqref{eq:NNA} is not simply the identity of a 4D absolute theory, corresponding to the fact that there is no direct counterpart of equation \eqref{eq:functor} upon performing a double gauging involving two distinct subalgebras of the gaugeable algebra characterising the mother SymTFT. Given that the latter is basically equivalent to the existence of a well-defined partition function for the resulting absolute theory, we claim that, if might be able to define a counterpart of the functor \eqref{eq:functor} for the case depicted at the centre and RHS of figure \ref{fig:newidentity}, by composing the following chain of functors

\begin{figure}[ht!]    
\begin{center}   
\includegraphics[scale=1.25]{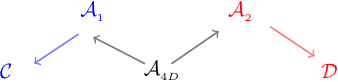}  
\caption{\small A qualitative diagram depicting the counterpart of the fiber functor's domain for the case of figure \ref{fig:newidentity}. The black arrors indicate the choice of different gaugeable subalgebras, leading to different 4D absolute theories, labelled by 2-categories ${\cal C,D}$.}   
\label{fig:3fibers}   
\end{center}   
\end{figure}  
where ${\cal A}_{_{4D}}$ is the gaugeable algebra leading to a single 4D absolute theory, whereas ${\cal A}_{_1}$ and ${\cal A}_{_2}$ correspond to different subalgebras of ${\cal A}_{_{4D}}$ leading to different 2-categories of underlying charges associated to the gauged symmetry characterising the two distinct 4D absolute theories. This is current work in progress by the same author, and we plan to report about it in a forthcoming work. The main message we want to convey with this is that the fiber functor that can be defined for the configuration at the centre and RHS of figure \ref{fig:newidentity} leads to the partition function of a 3D theory. We plan to report about this in due course, \cite{VP}.

\subsection{Quantum dimension from relative gaugeable algebra}  \label{sec:2.4}  

In the concluding part of this section, we make use of the tools outlined in the previous parts of the present treatment, leading to the proposal of a probe quantity enabling to distinguish whether the defect interpolating in between class ${\cal S}$ theories is intrinsically or non-intrinsically non-invertible. The starting point will be the setting of section \ref{sec:FMT}, where we revised the gauging-by-gauging procedure, implemented by flooding the topological order corresponding to the 2-category of Wilson surfaces living in the bulk SymTFT with a tensor network of an idempotent element, namely the norm element \eqref{eq:idemp}. 

\begin{figure}[ht!]    
\begin{center}   
\includegraphics[scale=1]{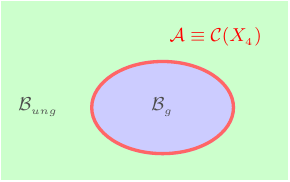}  
\caption{\small This picture shows the categories of anyons associated to the ungauged and gauged theories, denoted by ${\cal B}_{_{ung}}$ and ${\cal B}_{_g}$, respectively. Gauging-by-gauging can be visualised as the domain wall effectively separating two different topological phases.}   
\label{fig:gauging algebra}   
\end{center}   
\end{figure}

Practically, the two phases (prior and after gauging) can be thought of as arising from a phase transition in between two MTCs, ${\cal B}_{_{ung}}$ and ${\cal B}_{_g}$, corresponding to the ungauged and the gauged phases, respectively, \cite{Kong:2013aya}. Imposing natural physical requirenents, it is possible to derive a relation between the anyons in the ${\cal B}_{_{g}}$ phase and those in the ${\cal B}_{_{ung}}$ phase. The vacuum, or tensor unit ${\cal A}$ in ${\cal B}_{_g}$ is necessarily a connected commutative separable algebra in ${\cal B}_{_{ung}}$, and the category ${\cal B}_{_g}$ is equivalent to the category of local ${\cal A}$-modules as MTCs. This gauging produces a gapped domain wall (DW) with wall excitations given by the category of ${\cal A}$-modules in ${\cal B}_{_{ung}}$. The domain wall separating them, is, in turn a category, ${\cal A}\equiv{\cal C}(X_{_4})$, as shown in figure \ref{fig:gauging algebra}. 

For any pair of elements, $M, N$, 

\begin{equation}  
\text{Hom}_{_{{\cal B}_{_g}}}(M,N)\ \hookrightarrow \ \text{Hom}_{_{{\cal B}_{_{ung}}}}(M,N).
\end{equation}  

Assuming vacuum degeneracy being trivial, 

\begin{equation}  
\text{Hom}_{_{{\cal A}}}(\mathbb{1}_{_{\cal A}},\mathbb{1}_{_{\cal A}}) \ \simeq \ \mathbb{C},  
\end{equation}    
${\cal A}$ must be a UFC with a unique spherical structure. All objects in ${\cal A}$ should come from objects in ${\cal B}_{_{ung}}$, and should satisfy the following properties:  

\begin{enumerate}

\item ${\cal A}$ is a subcategory of ${\cal B}_{_{ung}}$   

\item \begin{equation}  
\text{Hom}_{_{{\cal A}}}(M,N)\ \hookrightarrow\   \text{Hom}_{_{{\cal B}_{_{ung}}}}(M,N)
\end{equation} 

\item    $\mathbb{1}_{_{ung}}$ should gauge into $B$, such that 

\begin{equation} 
\iota_{_B}:\ \mathbb{1}_{_{{\cal B}_{_{ung}}}}\ \rightarrow\ B\ \in\ {\cal B}_{_{ung}}  
\end{equation}   

$\mathbb{1}_{_{{\cal B}_{_g}}}\equiv{\cal A}$ should fuse into the vacuum on the wall when moving ${\cal A}$ close to the wall, such that, in ${\cal B}_{_{ung}}$, the corresponding map reads

\begin{equation} 
\iota_{_A}^{^B}:\ A\ \rightarrow\ B  
\end{equation}  

\end{enumerate}

What we have just said can be succinctly re-expressed as follows  

\begin{equation}  
\boxed{\ \ \mathbb{1}_{_{{\cal B}_{_g}}}\ \equiv\ {\cal A}\ \subset\ {\cal B}_{_{ung}} \color{white}\bigg]\ \ }. 
\label{eq:identityoperator}
\end{equation}

After what we have just said, the key point to focus on is the fact that the gauging-by-gauging procedure, the identity operator changes, in the sense that, from being a simple element in the category ${\cal B}_{_{ung}}$, it is nontrivial in the category of the gauged theory, ${\cal B}_{_g}$, since $\mathbb{1}_{_{{\cal B}_{_g}}}\equiv{\cal A}$. 

The analogy we have drawn so far with the example shown in figure \eqref{eq:identityoperator} taken from \cite{Kong:2013aya}, corresponds to the case of a single gauging of the 5D SymTFT, leading to the LHS of our figure \ref{fig:newidentity}. For the case involving a double algebraic gauging, instead, such prescription needs to be suitably adapted. The motivation still follows the arguments presented in \cite{Kong:2013aya}. We now argue why this is indeed the case specifying to the example described in section \ref{sec:4dfrom6d} and \ref{sec:2}.

Recall that ${\cal C}(X_{_4})$ is the operator implementing the outer automorphism on the category of defects defining the 5D SymTFT, and corresponds to the idempotent object that is constituted of the algebra elements that are flooding the SymTFT in order to implement the gauging. Denoting by ${\cal C}_{_g}, {\cal C}_{_{ung}}$ the 2-category of defects characterising the 4D absolute theories, if the two phases are related by a single bulk gauging-by-gauging, it follows that

\begin{equation}  
\text{dim}_{_{{\cal C}_{_g}}}\ \mathbb{1}_{_{{\cal C}_{_g}}}\ \equiv\ \frac{\ \text{dim}_{_{{\cal C}_{_{ung}}}} {\cal A}_{_{4D}}\ }{\text{dim}_{_{{\cal C}_{_{ung}}}} {\cal A}_{_{4D}}}\ \equiv\ 1.  
\label{eq:BBungg5}   
\end{equation}

But this is only from the point of view of the gauged theory, ${\cal C}_{_g}$. On the other hand, ${\cal A}_{_{4D}}$ is a graded object in ${\cal C}_{_{ung}}$, and therefore 

\begin{equation}  
\text{dim}_{_{{\cal C}_{_{ung}}}}\ {\cal A}_{_{4D}}\ \neq\ 1.     
\label{eq:BBungg6}   
\end{equation}

However, our aim is that of describing a composite configuration for the absolute theory of the kind depicted on the LHS of figure \ref{fig:gauging algebra2}, where, as explained in sections \ref{sec:4dfrom6d} and \ref{sec:2}, both phases really descend form the gauging of different subalgebras within the gaugeable algebra of the original ungauged SymTFT\footnote{This is our proposal on the basis of \cite{TJF,Kong:2013aya}.}. The different 4D absolute gauge theories are interpolated by a non-invertible defect defined as 

\begin{equation} 
{\cal N}({\cal M}_{_3})\ \overset{def.}{\equiv}\ {\cal D}_{_3}\ {\cal A}_{_{\epsilon\rho}} 
\label{eq:noninv}   
\end{equation}   
where ${\cal A}_{_{\epsilon\rho}}$ plays the role of a relative gaugeable algebra that cannot be identified with the vacuum on either side.

\begin{figure}[ht!]    
\begin{center}   
\includegraphics[scale=1]{fnf.pdf}  
\ \ \ \ \ \ \ \ \ \ \ \ \ \ \ \ \ \ \ \ \ \ \ \ \ \ \ 
\includegraphics[scale=1]{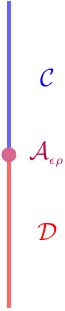}  
\caption{\small  Composite 4D absolute theories separated by an intrinsic non-invertible defect on the LHS. Rephrased in terms of the higher-categorical structure underlying the resulting theory, the algebra dressing the interpolating defects ${\cal N}_{_3}$ is a homomorphism between the objects of a 2-category.}   
\label{fig:gauging algebra22}   
\end{center}   
\end{figure}

Consistency of the configuration associated to the LHS of figure \ref{fig:gauging algebra22}, is mapped to a statement of Witt equivalence between the underlying 2-categories depicted on the RHS of the same figure, \cite{Kong:2013aya},

\begin{equation} 
{\cal C}\ \boxtimes\ \bar{\cal D}\ \simeq\ \mathfrak{Z}({\cal {\cal A}_{_{\epsilon\rho}}}), 
\label{eq:inclusion}   
\end{equation}   
where ${\cal A}_{_{\epsilon\rho}}$ denotes the relative gaugeable algebra featuring in \eqref{eq:noninv}. 

The relative gaugeable algebra, ${\cal A}_{_{\epsilon\rho}}$, constitutes a fusion tensor category\footnote{With ${\cal A}_{_1}$ and ${\cal A}_{_0}$ denoting the object and the morphisms, respectively.} defining morphisms between any pair of elements $L, L^{\prime}\ \in\ {\cal C,D}$.

\begin{equation}    
{\cal A}_{_{\epsilon\rho}}^{^{L,L^{^{\prime}}}} \ \overset{def.}{=}\ \left\{{\cal A}_{_1}^{^{L,L^{^{\prime}}}}, {\cal A}_{_0}^{^{L,L^{^{\prime}}}}\ \right\}, 
\end{equation}
whose total quantum dimension can be compared to that of the identity, i.e. of the algebra that has beengauged to achieve the reference gauge theory, with 2-category ${\cal C}$.


\begin{figure}[ht!]    
\begin{center}   
\includegraphics[scale=1]{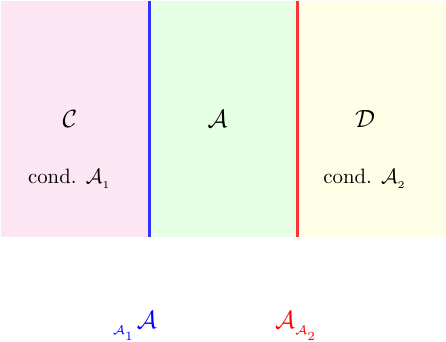}  
\caption{\small The algebra for the underlying category, denoted by ${\cal A}$ can be partiallygauged, under suitable choice of subalgebras ${\cal A}_{_1}, {\cal A}_{_2}\ \subset\ {\cal A}$, such that they lead to different gauge theories, whose 2-categories are labelled by ${\cal C, D}$.}   
\label{fig:gauging algebra1}   
\end{center}   
\end{figure}

\begin{figure}[ht!]    
\begin{center}   
\includegraphics[scale=1]{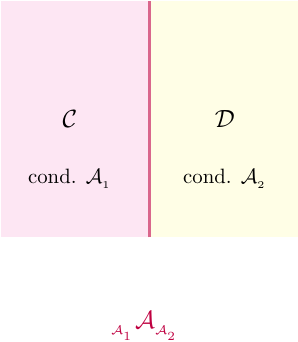}  
\caption{\small The composite system \eqref{fig:gauging algebra1} can be thought of as describing a thickened domain wall, $_{_{{\cal A}_{_1}}}{\cal A}_{_{{\cal A}_{_2}}}\ \overset{def.}{\equiv}\ {\cal A}_{_1}\ \otimes_{_{\cal A}}\ {\cal A}_{_2}$.}   
\label{fig:gauging algebra2}   
\end{center}   
\end{figure}

From the point of view of the absolute 4D theory, the non-invertible defect can be thought of as a thickened domain wall obtained by gauging two different subalgebras within the original gauging algebra ${\cal A}$, as shown in figure \ref{fig:gauging algebra2}. Given that the target absolute theory we want to achieve is that of figure \ref{fig:gauging algebra1}, then the corresponding 2-categories of defects can be obtained starting from 

\begin{equation}   
{\cal A}\ \equiv\ {\cal D}\ \boxtimes\ \mathfrak{Z}({\cal E}),  
\end{equation}

\begin{equation}  
G:\ {\cal C}\ \boxtimes\ \mathfrak{Z}({\cal D})\ \simeq\ {\cal D}\ \boxtimes\ \mathfrak{Z}({\cal E}), 
\end{equation}  
and the  fiber functors 

\begin{equation}  
{\cal F}_{_{\cal E}}\ :\ \mathfrak{Z}({\cal E})\ \rightarrow\ {\cal E}  
\ \ \ ,\ \ \ 
{\cal F}_{_{\cal D}}\ :\ \mathfrak{Z}({\cal D})\ \rightarrow\ {\cal D}   
\end{equation}   
with corresponding duals ${\cal F}_{_{\cal E}}^{^V}, {\cal F}_{_{\cal D}}^{^V}$.






The fact that ${\cal N}({\cal M}_{_{3}})$ is intrinsically non-invertible corresponds to saying that the outer automorphism implemented by the action of $F$ cannot be undone by a combination of modular transformations, as described in section \ref{sec:2}. In particular, ${\cal A}_{_{\epsilon\rho}}$ is different from the identity on either side of the defect, and, therefore, ${\cal D}_{_{{\cal A}_{_{\epsilon\rho}}}}\neq {\cal D}_{_{\mathbb{1}}}$. 

\section*{Proof of our main result}    

The concluding part of this section outlines the identification of a quantity we wish to propose as a probe for asserting whether a given non-invertible defect is either intrinsic or non-intrinsic. In doing so, we will be relying on most of the tools outlined in the previous parts of the present work.

Take ${\cal B}$ to be a semisimple abelian category over $\mathbb{C}$; assuming it is a rigid braided tensor category. $I$ is the set of isomorphism classes of irreducible objects in ${\cal B}$, with representatives $V_{_i}, \ \forall i\in I$. Assuming the spaces of morphisms are finite-dimensional. $\mathbb 1\in {\cal B}$ is assumed to be a simple object corresponding to the labelling $i\equiv0\in I\ \Rightarrow\ V_{_0}\overset{def.}{\equiv}\mathbb{1}$. A ${\cal B}$-algebra is an object ${\cal A}\in{\cal C}$ with morphisms $\mu:\ {\cal A}\otimes{\cal A}\rightarrow{\cal A}$ and $\iota_{_{{\cal A}}}:\mathbb{1}\hookrightarrow{\cal A}$, with 

\begin{equation} 
\text{dim \ Hom}_{_{{\cal B}}}(\mathbb{1},{\cal A})\ \equiv\ 1.  
\end{equation}

Given ${\cal B}$ and a ${\cal B}$-algebra ${\cal A}$, Rep${\cal A}$ is defined as pairs $(V,\mu_{_{V}})$, where $V\in{\cal B}$ and $\mu_{_{V}}:{\cal A}\otimes V\rightarrow V$ morphisms in ${\cal B}$. 

For example, if $G$ is a finite group and ${\cal B}$ is the category of finite-dimensional complex representations of $G$, semisimple ${\cal B}$-algebras correspond to different semigroups in $G$ (this is what we need to implement gauging-by-gauging in the Freed-Moore-Teleman setup). 

Here, in Rep${\cal A}$, ${\cal A}$ is the unit object. For the case where there are two different absolute theories separated by an intrinsically non-invertible defect, the setup of joint absolute theories results from a double gauging procedure \cite{Kong:2013aya} and it makes sense to consider the composite object $X\otimes_{_{{\cal A}}}Y$ with ${\cal A}$ the gaugeable algebra in ${\cal C}$; so, we have that $X,Y\in$ Rep${\cal A}$. But, at the same time, $X$ and $Y$ are also elements of two different subalgebras within ${\cal A}$ corresponding to two different gauging-by-gaugings. E.g. $X\in$ Rep${\cal A}_{_1}$ and $Y\in$ Rep${\cal A}_{_2}$, with ${\cal A}{_{_1}}, {\cal A}_{_{2}}\subset{\cal A}$. In turn, ${\cal A}_{_1}$ is the unit element in Rep${\cal A}_{_1}$ and ${\cal A}_{_2}$ id the unite element in Rep${\cal A}_{_2}$. 

Hence, if we consider the original mother algebra ${\cal A}$ with its characteristic elementary objects $V_{_i}$, and assuming every one of them has multiplicity $1$ within ${\cal A}$, we get 

\begin{equation}  
\text{dim}_{_{\cal A}}(X\otimes_{_{{\cal A}}}Y) \ \equiv\ \text{dim}_{_{\cal A}}(X)\ \text{dim}_{_{\cal A}}(Y)
\label{eq:first}
\end{equation}  
and 

\begin{equation}  
\text{dim}_{_{\cal A}}(X) \ \equiv\ \frac{\text{dim}_{_{\cal B}}({ X})}{\text{dim}_{_{\cal B}}({\cal A})}.   
\label{eq:second}
\end{equation} 

Furthermore, we have that

\begin{equation}  
\text{dim}_{_{\cal A}}(X\otimes_{_{\cal A}}Y) \ \equiv\ \frac{\text{dim}_{_{\cal B}}({ X})\ \text{dim}_{_{\cal B}}({ Y})}{\text{dim}_{_{\cal B}}({\cal A})}.   
\label{eq:third}
\end{equation}

Extending equation \eqref{eq:third} to the case in which $X, Y$ are replaced by the ${\cal A}_{_1}, {\cal A}_{_2}$ subalgebras themselves, we can look at the following ratio

\begin{equation}  
\frac{\text{dim}_{_{\cal B}}({\cal A}_{_1}\otimes_{_{\cal A}}{\cal A}_{_2})}{\text{dim}_{_{\cal B}}({\cal A}_{_1})}\ \equiv\ \frac{{\cal D}^{^{intr.}}_{_{_{{\cal A}_{_1}}}{\cal A}_{_{{\cal A}_{_2}}}}}{{\cal D}^{^{non-intr.}}_{{\cal A}_{_1}}},  
\label{eq:third1}
\end{equation} 
where on the RHS we identified the numerator and denominator with the quantum dimension of algebras associated to the intrinsic or non-intrinsic non-invertible defects separating absolute theories resulting from the Freed-Moore-Teleman construction discussed in the first part of section \ref{sec:FMT}. Making use of \eqref{eq:first} and \eqref{eq:second}, \eqref{eq:third1}, we get

\begin{equation}  
\frac{\text{dim}_{_{\cal A}}({\cal A}_{_1}\otimes_{_{\cal A}}{\cal A}_{_2})\ \text{dim}_{_{\cal B}}({\cal A})}{ \text{dim}_{_{\cal A}}({\cal A}_{_1})\ \text{dim}_{_{\cal B}}({\cal A})}\ \equiv\ \frac{\text{dim}_{_{\cal A}}({\cal A}_{_1})\ \text{dim}_{_{\cal A}}({\cal A}_{_2})}{\text{dim}_{_{\cal A}}({\cal A}_{_1})}\ \equiv\ \text{dim}_{_{\cal A}}({\cal A}_{_2}).      
\label{eq:third22}
\end{equation} 

Given that 

\begin{equation} 
\text{dim}_{_{\cal A}}({\cal A}_{_2})\ \overset{def.}{=}\ \sqrt{\sum_{_i}d_{_i}\ } 
\end{equation} 
where 

\begin{equation}  
d_{_i}\ \overset{def.}{=}\ \text{dim}_{_{\cal A}}V_{_i}\ \ >\ 1,   
\end{equation}
with $V_{_i}\in {\cal A}_{_2}$ being simple objects\footnote{I.e. a basis.} in ${\cal A}_{_2}$ (similar arguments hold for ${\cal A}_{_1}$, since both are assumed to be subalgebras in ${\cal A}$), it therefore follows that 

\begin{equation}  
\boxed{\ \ \ \frac{\text{dim}_{_{\cal B}}({\cal A}_{_1}\otimes_{_{\cal A}}{\cal A}_{_2})}{\text{dim}_{_{\cal B}}({\cal A}_{_1})}\ \equiv\ \frac{{\cal D}^{^{intr.}}_{_{_{_{{\cal A}_{_1}}}{\cal A}_{_{{\cal A}_{_2}}}}}}{{\cal D}^{^{non-intr.}}_{{\cal A}_{_1}}}\ >\ 1\color{white}\bigg]\ \ }.  
\label{eq:third2}
\end{equation}

For the configuration of interest, namely two different absolute theories obtained by gauging-by-gauging of two different subalgebras ${\cal A}_{_1}, {\cal A}_{_2}\subset{\cal A}$, separated by an intrinsic non-invertible defect, the numerator in \eqref{eq:third2} corresponds to the quantum dimension of the nontrivial fusion of an intrinsic non-invertible defect and its orientation reverse, whereas the denominator corresponds to the quantum dimension of the algebra that has beengauged in the SymTFT to obtain a specific absolute theory. 
We are therefore led to conclude that the total quantum dimension of the relative gaugeable algebra ${\cal A}_{_{\epsilon\rho}}$ is greater w.r.t. that of  the gauged algebra leading to the single gauged phase, ${\cal C}$, and we propose \eqref{eq:third2} as a probe quantity signalling whether a non-invertible defect is either intrinsic or non-intrinsic.

For the specific case of class ${\cal S}$ theories, the algebras featuring in \eqref{eq:third2} explicitly read

\begin{equation} 
{\cal N}({\cal M}_{_{3}})\ \times\ \bar {\cal N}({\cal M}_{_{3}}) 
\equiv{\cal A}_{_{\epsilon\rho}}    
\label{eq:NN}
\end{equation} 

\begin{equation} 
{\cal D}_{_3}\ \times\ \bar {\cal D}_{_3} \equiv\ \mathbb{1}\ 
\equiv\ {\cal A}_{_1}. 
\label{eq:DD}
\end{equation}

Gathering these results together, we can rewrite equation \eqref{eq:third2} in terms of \eqref{eq:NN} and \eqref{eq:DD}, leading to the following result

\begin{equation} 
\begin{aligned}    
\boxed{\ \ \ \frac{<{\cal N}({\cal M}_{_{3}})\ \times\ \bar {\cal N}({\cal M}_{_{3}}) >_{_{{\cal B}}}}{<{\cal D}_{3}\ \times\ \bar{\cal D}_{3}>_{_{{\cal B}}}}\  \equiv\ \frac{ <{\cal A}_{_{\epsilon\rho}} >_{_{{\cal B}}} }{ \ <\mathbb{1}_{_{{\cal C}}} >_{_{{\cal B}}} \ }\ \equiv\ \left(\frac{{\cal D}^{^{intr.}}_{{\cal A}_{_{\epsilon\rho}}}}{{\cal D}^{^{non-intr.}}_{{\cal A}_{_1}}} \right)_{_{X_{4} }}\ >\ 1\color{white}\bigg]\ \ },      
\label{eq:4DNN3}
\end{aligned}  
\end{equation} 
with the last relation indicating that such ratio can be used as a parameter probing whether the two gauge theories are separated by an intrinsic non-invertible symmetry. Consequently, the inequality 

\begin{equation}  
\boxed{\ \ \ {\cal D}_{{\cal A}_{_{\epsilon\rho}}{(X_{_{4}})}}^{^{intr.}} \ >  \ {\cal D}_{{\cal A}_{_{\epsilon\rho}}{(X_{_{4}})}}^{^{non-intr. }}\color{white}\bigg]\color{black}\ \  }   
\label{eq:BFE}  
\end{equation} 
implies that the fusion category associated to intrinsic non-invertible defects is characterised by a greater quantum dimension, implying such configuration (arising from double gauging) is able to store a higher number of degrees of freedom w.r.t. the case in which a unique gauging is taking place. \footnote{In \eqref{eq:BFE}  we explicitly used the pedix ${\cal A}_{_{\epsilon\rho}}$ on both sides to indicate that the defect we are considering is separating absolute theories with different choices of boundary conditions on the topological boundary, according to the Freed-Moore-Teleman prescription. The difference in between the intrinsic and non-intrinsic case depends on whether such boundary conditions are isomorphic to each other or not.}

\section*{Key points}  

\begin{enumerate}  

\item For class ${\cal S}$ theories, ${\cal D}_{{\cal A}_{_{\epsilon\rho}}}$ enables to probe whether a defect is intrinsically or non-intrinsically non-invertible.  

\item ${\cal D}_{{\cal A}_{_{\epsilon\rho}}{(X_{_{4}})}}^{^{intr.}} \ >  \ {\cal D}_{{\cal A}_{_{\epsilon\rho}}{(X_{_{4}})}}^{^{non-intr. }}$, and therefore the former is a theory equipped with a higher-categorical structure enabling to store more d.o.f. in certain superselection sectors. 

\item For theories descending from 6D ${\cal N}=(2,0)$ SCFTs, if the gauging procedure leads to the presence of intrinsic non-invertible defects, the total quantum dimension of the algebra implementing the non-invertibility of such defects exhibits an increase in its total quantum dimension, signalling the relative gaugeable algebra ${\cal A}_{_{\epsilon\rho}}$ is characterised by superselection sectors admitting a richer structure w.r.t.  the gauged algebra leading to a single absolute 4D gauged theory.

\end{enumerate}

\subsection{Examples}  

The main result stated in Section \ref{sec:2.4} has a very broad applicability, that was discussed by the same author in another work, \cite{VP}. The breadth of this result stems from the fact that we can make use of tools such as factorisation homology and full-dualisability in the context of cobordism theory to map class-${\cal S}$ theory settings to two-dimensional topological orders, by making use of the Alday-Gaiotto-Tachikawa correspondence. We refer the interested reader to \cite{VP} for a more detailed analysis in this regard.

\section{Conclusions and Outlook}    

Key advancements towards achieving a rigorous mathematical formulation of QFTs 
strongly rely upon a higher-categorical prescription, where the symmetries of a QFT can be thought of as defects living in different categories connected by functors.  

In the present work we identify a criterion to distinguish between intrinsic and non-intrinsic non-invertible symmetries arising from the categorical structure of 6D ${\cal N}=(2,0)$ SCFTs. In particular we were able to derive a relation in terms of the quantum dimension of the relative gauging algebra implementing the gauging in the bulk SymTFT and leading to absolute 4D theories interpolated by intrinsic non-invertible defects. In doing so, we relied upon the description in terms of relative field theories, as described in \cite{Bashmakov:2022uek} making use of the Freed-Moore-Teleman setup, \cite{Freed:2012bs,Freed:2022qnc,Freed:2022iao}, together with the gauging-by-gauging prescription, \cite{TJF}. For the intrinsic non-invertible case, multiplicity for superselection sectors of the relative gaugeable algebra is greater w.r.t. the non-intrinsic case, thereby signalling the possibility for additional d.o.f. to be stored in certain superselection sectors of the resulting composite absolute theory. 

Our results extend arguments proposed in \cite{Kaidi:2022cpf}, where the authors also proposed a way of distinguishing intrinsic from non-intrinsic non-invertibility in 2D by means of the quantum dimension of the non-invertible defect, which in terms of the underlying higher-categorical structure of the composite absolute 4D theory, plays the role of a fusion tensor category defining the morphisms between objects of the 2-categories, with the latter being the defects charged under the gauged symmetry. 

In our analysis, we encountered two major setups built upon the notion of higher-categories, namely symmetry TFTs (SymTFT) and topological orders (TO). Plenty of effort has been made towards building a correspondence in between the two, most recently in \cite{Ji:2019eqo}. Our findings provide further support towards strengthening the connection between the two description. 

En-passing, we commented on a proposal regarding the realisation of such composite absolute theories and the definition of a fiber functor intrinsically related to the notion of a partition function for a 3D theory rather than a 4D theory, which is current work in progress by the same author, \cite{VP}.  

We conclude by stressing that our analysis is mostly motivated by furthering the understanding of  6D ${\cal N}=(2,0)$ SCFTs, from, both, a mathematical and physical point of view. To what extent these findings might be mapped to other setups is currently under investigation\footnote{Among possible setups one could be interested in analysing are the models addressed in, e.g. \cite{Heckman:2022muc,Dierigl:2022reg}.}. In an upcoming work \cite{VP}, we reported a more detailed analysis in this regard, by making use of tools such as factorisation homology and full-dualisability in the context of cobordism theory, with suitable adaptation to class-${\cal S}$ by means of the AGT correspondence. We expect further investigation in terms of cobordisms could lead to potential further understanding of singularity theories and moduli spaces of varieties arising in algebraic geometry settings, which we aim to address in future work.

\section*{Acknowledgements}

The author wishes to thank Tudor Dimofte, I$\tilde{\text{n}}$aki Garc\'ia-Etxebarria, David Jordan, Fernando Quevedo and Constantin Teleman for very useful discussions and questions raised. A special acknowledgement is in order to the Hodge Institute, where this work was completed and first presented. She also acknowledges Gregory Moore and Nati Seiberg for insightful comments and questions raised while presenting this work at NHETC Rutgers in September 2023. This work is partially supported by an STFC scholarship through DAMTP.

\appendix


\begin{thebibliography}{99}   



\bibitem{Witten:1995zh}
E.~Witten,
\emph{Some comments on string dynamics},  
[arXiv:hep-th/9507121 [hep-th]], 
https://doi.org/10.48550/arXiv.hep-th/9507121
.     





\bibitem{Witten:2007ct}
E.~Witten,
\emph{Conformal Field Theory In Four And Six Dimensions},   
[arXiv:0712.0157 [math.RT]], 
https://doi.org/10.48550/arXiv.0712.0157
.



\bibitem{Chacaltana:2012zy}
O.~Chacaltana, J.~Distler and Y.~Tachikawa,
\emph{Nilpotent orbits and codimension-two defects of 6d N=(2,0) theories},   
Int. J. Mod. Phys. A \textbf{28} (2013), 1340006
doi:10.1142/S0217751X1340006X
[arXiv:1203.2930 [hep-th]].    

\bibitem{Tachikawa:2013hya}
Y.~Tachikawa,
\emph{On the 6d origin of discrete additional data of 4d gauge theories},   
JHEP \textbf{05} (2014), 020
doi:10.1007/JHEP05(2014)020
[arXiv:1309.0697 [hep-th]].


\bibitem{Moore}
G.~Moore, 
\emph{Applications of the six-dimensional (2,0) theories to Physical Mathematics}, Felix Klein lectures Bonn (2012). 


\bibitem{Witten:2009at}
E.~Witten,
\emph{Geometric Langlands From Six Dimensions},    
[arXiv:0905.2720 [hep-th]], 
https://doi.org/10.48550/arXiv.0905.2720
.


\bibitem{DBZ}
D.~Ben-Zvi, \emph{Theory X
 and Geometric Representation Theory}, talks at Mathematical Aspects of Six-Dimensional Quantum Field Theories IHES 2014, notes by Qiaochu Yuan.








\bibitem{Bah:2022wot}
I.~Bah, D.~S.~Freed, G.~W.~Moore, N.~Nekrasov, S.~S.~Razamat and S.~Schafer-Nameki,
\emph{A Panorama Of Physical Mathematics c. 2022},   
[arXiv:2211.04467 [hep-th]], 
https://doi.org/10.48550/arXiv.2211.04467 .


\bibitem{Moore:1988qv}
G.~W.~Moore and N.~Seiberg,
\emph{Classical and Quantum Conformal Field Theory},   
Commun. Math. Phys. \textbf{123} (1989), 177
doi:10.1007/BF01238857.

\bibitem{Moore:1988ss}
G.~W.~Moore and N.~Seiberg,
\emph{Naturality in Conformal Field Theory},
Nucl. Phys. B \textbf{313} (1989), 16-40
doi:10.1016/0550-3213(89)90511-7.


\bibitem{Verlinde:1988sn}
E.~P.~Verlinde,
\emph{Fusion Rules and Modular Transformations in 2D Conformal Field Theory}, 
Nucl. Phys. B \textbf{300} (1988), 360-376
doi:10.1016/0550-3213(88)90603-7. 

\bibitem{Gaiotto:2014kfa}
D.~Gaiotto, A.~Kapustin, N.~Seiberg and B.~Willett,
\emph{Generalized Global Symmetries}, 
JHEP \textbf{02} (2015), 172
doi:10.1007/JHEP02(2015)172
[arXiv:1412.5148 [hep-th]].

\bibitem{Gaiotto:2020iye}
D.~Gaiotto and J.~Kulp,
\emph{Orbifold groupoids}, 
JHEP \textbf{02} (2021), 132
doi:10.1007/JHEP02(2021)132
[arXiv:2008.05960 [hep-th]].


\bibitem{Kong:2020cie}
L.~Kong, T.~Lan, X.~G.~Wen, Z.~H.~Zhang and H.~Zheng,
\emph{Algebraic higher symmetry and categorical symmetry -- a holographic and entanglement view of symmetry},   
Phys. Rev. Res. \textbf{2} (2020) no.4, 043086
doi:10.1103/PhysRevResearch.2.043086
[arXiv:2005.14178 [cond-mat.str-el]].

\bibitem{Kong:2022cpy}
L.~Kong and Z.~H.~Zhang,
\emph{An invitation to topological orders and category theory}, 
[arXiv:2205.05565 [cond-mat.str-el]], 
https://doi.org/10.48550/arXiv.2205.05565
.

\bibitem{Kong:2019byq}
L.~Kong and H.~Zheng,
\emph{A mathematical theory of gapless edges of 2d topological orders. Part I}, 
JHEP \textbf{02} (2020), 150
doi:10.1007/JHEP02(2020)150
[arXiv:1905.04924 [cond-mat.str-el]].

\bibitem{Kong:2019cuu}
L.~Kong and H.~Zheng,
\emph{A mathematical theory of gapless edges of 2d topological orders. Part II},  
Nucl. Phys. B \textbf{966} (2021), 115384
doi:10.1016/j.nuclphysb.2021.115384
[arXiv:1912.01760 [cond-mat.str-el]].


\bibitem{Kong:lastbutone}
L.~Kong and H.~Zheng,
\emph{Categorical computation}, 
Front. Phys. 18(2), 21302 (2023)
[arXiv:2102.04814v2[quant-ph]], 
https://doi.org/10.1007/s11467-022-1251-5.



\bibitem{Gaiotto:2019xmp}
D.~Gaiotto and T.~Johnson-Freyd,
\emph{Condensations in higher categories}, 
[arXiv:1905.09566 [math.CT]], https://doi.org/10.48550/arXiv.1905.09566
.

\bibitem{Johnson-Freyd:2021tbq}
T.~Johnson-Freyd and M.~Yu,
\emph{Topological Orders in (4+1)-Dimensions},   
SciPost Phys. \textbf{13} (2022) no.3, 068
doi:10.21468/SciPostPhys.13.3.068
[arXiv:2104.04534 [hep-th]].



\bibitem{Johnson-Freyd:2020usu}
T.~Johnson-Freyd,
\emph{On the Classification of Topological Orders},   
Commun. Math. Phys. \textbf{393} (2022) no.2, 989-1033
doi:10.1007/s00220-022-04380-3
[arXiv:2003.06663 [math.CT]].


\bibitem{MYu}
T.~D.~D\'ecoppet and M.~Yu,
\emph{Gauging noninvertible defects: a 2-categorical perspective},
Lett. Math. Phys. \textbf{113} (2023) no.2, 36
doi:10.1007/s11005-023-01655-1
[arXiv:2211.08436 [math.CT]].




\bibitem{Freed:2012bs}
D.~S.~Freed and C.~Teleman,
\emph{Relative quantum field theory},    
Commun. Math. Phys. \textbf{326} (2014), 459-476
doi:10.1007/s00220-013-1880-1
[arXiv:1212.1692 [hep-th]].






\bibitem{Freed:2022qnc}
D.~S.~Freed, G.~W.~Moore and C.~Teleman,
\emph{Topological symmetry in quantum field theory},   
[arXiv:2209.07471 [hep-th]], 
https://doi.org/10.48550/arXiv.2209.07471
.     

\bibitem{Freed:2022iao}
D.~S.~Freed,
\emph{Introduction to topological symmetry in QFT},    
[arXiv:2212.00195 [hep-th]], 
https://doi.org/10.48550/arXiv.2212.00195
.   










\bibitem{Bashmakov:2022jtl}
V.~Bashmakov, M.~Del Zotto and A.~Hasan,
\emph{On the 6d Origin of Non-invertible Symmetries in 4d},   
[arXiv:2206.07073 [hep-th]], 
https://doi.org/10.48550/arXiv.2206.07073
.      

\bibitem{Bashmakov:2022uek}
V.~Bashmakov, M.~Del Zotto, A.~Hasan and J.~Kaidi,
\emph{Non-invertible Symmetries of Class $\mathcal{S}$ Theories},   
[arXiv:2211.05138 [hep-th]], 
https://doi.org/10.1007/JHEP05.   

\bibitem{Bhardwaj:2022kot}
L.~Bhardwaj, S.~Schafer-Nameki and A.~Tiwari,
\emph{Unifying Constructions of Non-Invertible Symmetries},  
[arXiv:2212.06159 [hep-th]], 
https://doi.org/10.21468/SciPostPhys.15.3.122
.
 



    

\bibitem{Bhardwaj:2022yxj}
L.~Bhardwaj, L.~E.~Bottini, S.~Schafer-Nameki and A.~Tiwari,
\emph{Non-Invertible Higher-Categorical Symmetries}, 
[arXiv:2204.06564 [hep-th]], 
https://doi.org/10.21468/SciPostPhys.14.1.007.





     


\bibitem{Kaidi:2021xfk}
J.~Kaidi, K.~Ohmori and Y.~Zheng,
\emph{Kramers-Wannier-like Duality Defects in (3+1)D Gauge Theories}, 
Phys. Rev. Lett. \textbf{128} (2022) no.11, 111601
doi:10.1103/PhysRevLett.128.111601
[arXiv:2111.01141 [hep-th]].

\bibitem{Kaidi:2022uux}
J.~Kaidi, G.~Zafrir and Y.~Zheng,
\emph{Non-invertible symmetries of $ \mathcal{N} $ = 4 SYM and twisted compactification}, 
JHEP \textbf{08} (2022), 053
doi:10.1007/JHEP08(2022)053
[arXiv:2205.01104 [hep-th]].  

\bibitem{Kaidi:2022cpf}
J.~Kaidi, K.~Ohmori and Y.~Zheng,
\emph{Symmetry TFTs for Non-Invertible Defects},  
[arXiv:2209.11062 [hep-th]], 
https://doi.org/10.48550/arXiv.2209.11062
.  


\bibitem{Burbano:2021loy}
I.~M.~Burbano, J.~Kulp and J.~Neuser,
\emph{Duality defects in E$_{8}$},   
JHEP \textbf{10} (2022), 186
doi:10.1007/JHEP10(2022)187
[arXiv:2112.14323 [hep-th]].


\bibitem{Choi:2022zal}
Y.~Choi, C.~Cordova, P.~S.~Hsin, H.~T.~Lam and S.~H.~Shao,
\emph{Non-invertible gauging, Duality, and Triality Defects in 3+1 Dimensions},    
[arXiv:2204.09025 [hep-th]], 
https://doi.org/10.1103/PhysRevD.105.125016.   



\bibitem{Choi:2021kmx}
Y.~Choi, C.~Cordova, P.~S.~Hsin, H.~T.~Lam and S.~H.~Shao,
\emph{Noninvertible duality defects in 3+1 dimensions},   
Phys. Rev. D \textbf{105} (2022) no.12, 125016
doi:10.1103/PhysRevD.105.125016
[arXiv:2111.01139 [hep-th]].  

\bibitem{Antinucci:2022vyk}
A.~Antinucci, F.~Benini, C.~Copetti, G.~Galati and G.~Rizi,
\emph{The holography of non-invertible self-duality symmetries}, 
[arXiv:2210.09146 [hep-th]], 
https://doi.org/10.48550/arXiv.2210.09146
.

\bibitem{Antinucci:2022cdi}
A.~Antinucci, C.~Copetti, G.~Galati and G.~Rizi,
\emph{''Zoology'' of non-invertible duality defects: the view from class $\mathcal{S}$},    
[arXiv:2212.09549 [hep-th]], 
https://doi.org/10.48550/arXiv.2212.09549
.








\bibitem{TJF}    
T.~Johnson-Freyd,    
\emph{Operators and higher categories in quantum field theory}, Lecture series.   


\bibitem{Kong:2013aya}
L.~Kong,
\emph{Anyon gauging and tensor categories},   
Nucl. Phys. B \textbf{886} (2014), 436-482
doi:10.1016/j.nuclphysb.2014.07.003
[arXiv:1307.8244 [cond-mat.str-el]].


\bibitem{Yu:2021zmu}
M.~Yu,
\emph{Gauging Categorical Symmetries in 3d Topological Orders and Bulk Reconstruction},
[arXiv:2111.13697 [hep-th]], 
https://doi.org/10.48550/arXiv.2111.13697
.



\bibitem{Aharony:2013hda}
O.~Aharony, N.~Seiberg and Y.~Tachikawa,
\emph{Reading between the lines of four-dimensional gauge theories},
JHEP \textbf{08} (2013), 115
doi:10.1007/JHEP08(2013)115
[arXiv:1305.0318 [hep-th]].   




 










\bibitem{Ji:2019eqo}
W.~Ji and X.~G.~Wen,
\emph{Non-invertible anomalies and mapping-class-group transformation of anomalous partition functions}, 
Phys. Rev. Research. \textbf{1} (2019), 033054
doi:10.1103/PhysRevResearch.1.033054
[arXiv:1905.13279 [cond-mat.str-el]].
































\bibitem{Teo:2015xla}
J.~C.~Y.~Teo, T.~L.~Hughes and E.~Fradkin,
\emph{Theory of Twist Liquids: Gauging an Anyonic Symmetry},   
Annals Phys. \textbf{360} (2015), 349-445
doi:10.1016/j.aop.2015.05.012
[arXiv:1503.06812 [cond-mat.str-el]].



\bibitem{Barkeshli:2014cna}
M.~Barkeshli, P.~Bonderson, M.~Cheng and Z.~Wang,
\emph{Symmetry Fractionalization, Defects, and Gauging of Topological Phases},  
Phys. Rev. B \textbf{100} (2019) no.11, 115147
doi:10.1103/PhysRevB.100.115147
[arXiv:1410.4540 [cond-mat.str-el]].

  



\bibitem{VP}
V.~Pasquarella, \emph{Factorisation Homology for Class \cal S Theories},     	arXiv:2312.06760 [hep-th], 
https://doi.org/10.48550/arXiv.2312.06760
.


\bibitem{Heckman:2022muc}
J.~J.~Heckman, M.~H\"ubner, E.~Torres and H.~Y.~Zhang,
\emph{The Branes Behind Generalized Symmetry Operators}, 
Fortsch. Phys. \textbf{71} (2023) no.1, 2200180
doi:10.1002/prop.202200180
[arXiv:2209.03343 [hep-th]].


\bibitem{Dierigl:2022reg}
M.~Dierigl, J.~J.~Heckman, M.~Montero and E.~Torres,
\emph{IIB string theory explored: Reflection 7-branes}, 
Phys. Rev. D \textbf{107} (2023) no.8, 086015
doi:10.1103/PhysRevD.107.086015
[arXiv:2212.05077 [hep-th]].




\end{thebibliography}
\end{document}